\documentclass[journal]{IEEEtran}
\usepackage{subfigure}
\usepackage{bm}
\usepackage{multirow}
\ifCLASSINFOpdf
  \usepackage[pdftex]{graphicx}
  \graphicspath{{../pdf/}{../jpeg/}}
  \DeclareGraphicsExtensions{.pdf,.jpeg,.png}
\else
  \usepackage[dvips]{graphicx}
  \graphicspath{{../eps/}}
  \DeclareGraphicsExtensions{.eps}
\fi

\usepackage{cite}

\ifCLASSINFOpdf
\else
\fi

\usepackage{amsmath}

\usepackage{amssymb}
\usepackage{latexsym}
\usepackage{booktabs}

\begin{document}
%
\title{A Comprehensive Study on Temporal Modeling for Online Action Detection}
%
%
%

\author{Wen~Wang, Xiaojiang~Peng, Yu Qiao, Jian Cheng
\thanks{Wen wang and Jian Cheng are with the School of Information and Communication Engineering, University of Electronic Science and Technology of China, Chengdu, Sichuan, China, 611731.}
\thanks{Xiaojiang Peng and Yu Qiao are with ShenZhen Key Lab of Computer Vision and Pattern Recognition, SIAT-SenseTime Joint Lab, Shenzhen Institutes of Advanced Technology, Chinese Academy of Sciences; SIAT Branch, Shenzhen Institute of Artificial Intelligence and Robotics for Society.}
\thanks{This work was done when Wen Wang was intern at Shenzhen Institutes of Advanced Technology, Chinese Academy of Sciences.}
\thanks{Corresponding author: Xiaojiang Peng (xj.peng@siat.ac.cn) and Jian Cheng (chengjian@uestc.edu.cn)}}

%
%

\newcommand{\etal}{\textit{et al}. }
\newcommand{\ie}{\textit{i}.\textit{e}. }
\newcommand{\eg}{\textit{e}.\textit{g}. }

\markboth{Journal of \LaTeX\ Class Files,~Vol.~14, No.~8, August~2015}%
{Shell \MakeLowercase{\textit{et al.}}: Bare Demo of IEEEtran.cls for IEEE Journals}
%



\maketitle

\begin{abstract}
Online action detection (OAD) is a practical yet challenging task, which has attracted increasing attention in recent years. A typical OAD system mainly consists of three modules: a frame-level feature extractor which is usually based on pre-trained deep Convolutional Neural Networks (CNNs), a temporal modeling module, and an action classifier. Among them, the temporal modeling module is crucial which aggregates discriminative information from historical and current features. Though many temporal modeling methods have been developed for OAD and other topics, their effects are lack of investigation on OAD fairly. This paper aims to provide a comprehensive study on temporal modeling for OAD including four meta types of temporal modeling methods, \ie temporal pooling, temporal convolution, recurrent neural networks, and temporal attention, and uncover some good practices to produce a state-of-the-art OAD system. Many of them are explored in OAD for the first time, and extensively evaluated with various hyper parameters. Furthermore, based on our comprehensive study, we present several hybrid temporal modeling methods, which outperform the recent state-of-the-art methods  with sizable margins on THUMOS-14 and TVSeries.
\end{abstract}

\begin{IEEEkeywords}
Online action detection, temporal pooling, temporal convolution, recurrent neural network, temporal attention.
\end{IEEEkeywords}

%
\IEEEpeerreviewmaketitle

\section{Introduction}
\label{sec1}
Online action detection (OAD) is an important problem in computer vision, which has wide range of applications like visual surveillance, human-computer interaction, and intelligent robot navigation, etc. Different from traditional action recognition and offline action detection that intend to recognize actions from full videos, the goal of online action detection is to detect an action as it happens and ideally even before the action is fully completed. It is a very challenging problem due to the extra restriction on the usage of only historical and current information except for the difficulties of traditional action recognition in untrimmed video streams.

In general, there exist two OAD tasks, \ie spatial-temporal online action detection (ST OAD) and temporal online action detection. With online setting, the former aims to localize actors and recognize actions in space-time which is introduced in \cite{DBLP:conf/cvpr/SoomroIS16}, while the latter is to localize and recognize actions temporally only which is systematically introduced in \cite{De2016Online}. Our study mainly focuses on the temporal online action detection problem, and we ignore `temporal' for convenience in the rest.

\begin{figure}
\centering
\includegraphics[width=1.0\linewidth]{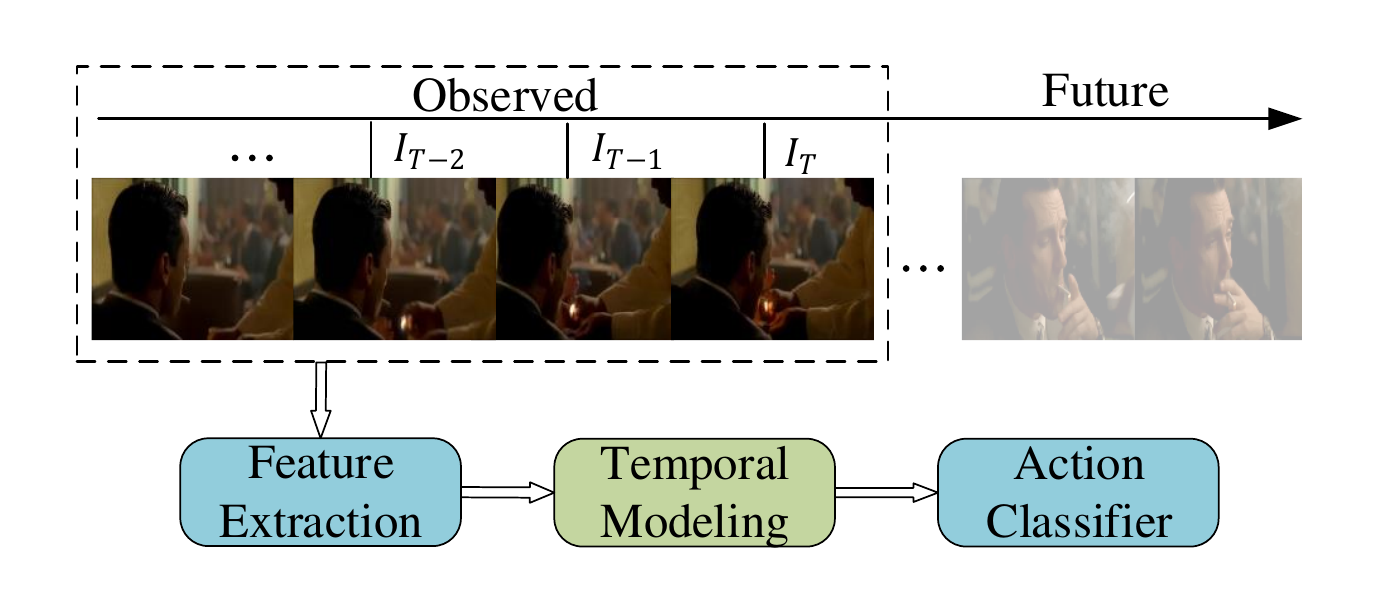}
  \caption{Online action detection aims to predict the ongoing action category from the historical and current frame information. A typical online action detection system is mainly composed of three parts: frame-level feature extraction, temporal modeling, and action classification.}
\label{fig_0}
\end{figure}

As illustrated in Fig.\ref{fig_0}, an online action detection (OAD) system mainly consists of three important parts: a frame-level feature extractor (\textit{e.g}. deep Convolutional Neural Network, CNN), a temporal modeling module to aggregate frame-level features, and an action classifier.
Recent works on online action detection mostly focus on the temporal modeling part, aiming to generate discriminative representations from the historical and current frame features. Inspired by the sequence modeling methods in other areas especially the Long Short-Term Memory recurrent network (LSTM)~\cite{DBLP:journals/neco/HochreiterS97}, various temporal modeling methods have been developed for online action detection recently. For example, Geest \etal \cite{De2016Online} provide a LSTM-based baseline which shows superiority to the single-frame CNN model.
Gao \etal~\cite{DBLP:conf/bmvc/GaoYN17a} propose a LSTM-based Reinforced Encoder-Decoder network for both action anticipation and online action detection. Geest \etal~\cite{De} propose a two-stream feedback network, where one stream focuses on the input interpretation and the other models temporal dependencies between actions. Xu \etal~\cite{Xu_2019_ICCV} utilize LSTM cell to model temporal context aiming to improve online action detection by adding prediction information into observed information.

Although the above LSTM-based temporal modeling methods have significantly boosted the performance on existing OAD datasets (\eg TVSeries \cite{De2016Online}, THUMOS-14 \cite{THUMOS14}), however, their superiority to other temporal models, \eg, naive temporal pooling, temporal convolution, and attention-based sequence models, is not discussed and remains unknown. Moreover, the fusion of different temporal models is also rarely investigated. To address these problems, we provide a fair comprehensive study on temporal modeling for online action detection in the following aspects.

\textbf{Exploration of temporal modeling methods}. We explore four popular types of temporal modeling methods with various hyper parameters to fairly illustrate their effects for online action detection. They are namely temporal pooling, temporal convolution, recurrent neural networks, and temporal attention models. Specifically, for temporal pooling, we evaluate \textit{average pooling} (AvgPool) and \textit{max pooling} (MaxPool) with various sequence lengths. For temporal convolution, we evaluate traditional \textit{temporal convolution} (TC), \textit{pyramid dilated temporal convolution} (PDC) \cite{LiGlobal}, and \textit{dilated causal convolution} (DCC) \cite{OordWaveNet}. For recurrent neural networks, we evaluate \textit{LSTM} and \textit{Gated Recurrent Unit} (GRU) with two output choices, \ie the last hidden state and the average hidden state. For temporal attention, we evaluate \textit{naive self-attention} (Naive-SA) with a linear fully-connected (FC) layer and Softmax function, \textit{nonlinear self-attention} (Nonlinear-SA) with a FC-tanh-FC-Softmax architecture, \textit{Non-local} block or standard self-attention with a skip connection, and \textit{Transformer} with the current frame as the query (Q) information.
It is worth noting that i) we try to keep the original names of these related methods in other topics though we make adaptions for online action detection, and ii) many of these methods are introduced into online action detection for the first time to the best of our knowledge, such as TC, PDC, DCC, Non-local, etc.
Overall, we extensively explore eleven individual temporal modeling methods with the off-the-shelf two-stream (TS) frame features.

\textbf{The fusion of temporal modeling methods}. Generally, these sequence-to-sequence methods, \eg  PDC and LSTM, can be further processed by aggregation methods to create a single representation like temporal pooling and temporal attention. Thus, we present several hybrid temporal modeling methods which combine different temporal modeling methods aiming to uncover the complementarity among them. Interestingly, we find that a simple fusion between dilated causal convolution and Transformer or LSTM improves the individual models significantly.

\textbf{Comparison with state of the arts.} We extensively compare our individual models and hybrid temporal models to existing baselines and recent state-of-the-art methods. Several hybrid temporal models outperform the best existing performance with a sizable margin on both TVSeries and THUMOS-14. Specifically, the fusion of dilated causal convolution and Transformer obtains \textbf{84.3}\% cAP on TVSeries, and the fusion of dilated causal convolution, LSTM, and Transformer achieves \textbf{48.6}\% mAP on THUMOS-14.

\section{Related Work}
Our study is related to several other action related tasks, namely action recognition, action anticipation, temporal action detection, spatial-temporal action detection. In this section, we first briefly overview these related tasks separately and then present the recent works on online action detection.

\textbf{Action recognition} is an important branch of video related research areas and has been extensively studied in the past decades. The existing methods are mainly developed for extracting discriminative action features from temporally complete action videos. These methods can be roughly categorized into hand-crafted feature based approaches and deep learning based approaches. Early methods such as Improved Dense Trajectory (IDT) mainly adopt hand-crafted features, such as HOF \cite{Laptev2008Learning}, HOG \cite{Laptev2008Learning} and MBH \cite{Wang2013Dense}. Recent studies demonstrate that action features can be learned by deep learning methods such as convolutional neural networks (CNN) and recurrent neural networks (RNN). For example, two-stream network \cite{Simonyan2014Two, Wang2015Towards} learns appearance and motion features based on RGB frame and optical flow field separately. RNNs, such as long short-term memory (LSTM) \cite{Graves1997Long} and gated recurrent unit (GRU) \cite{DBLP:journals/corr/ChoMBB14}, have been used to model long-term temporal correlations and motion information in videos, and generate video representation for action classification. Some recent works also try to model temporal information within a 2D-CNN instead of using 2D-CNN as static feature extractor, \eg both TSM \cite{DBLP:journals/corr/abs-1811-08383} and TAM \cite{NIPS2019_8498} propose an efficient approach to aggregate feature across frames inside the network.

Another type of action recognition approach is based on 3D CNNs, which are widely used for learning large-scale video datasets. C3D~\cite{Du2015Learning} is the first successful 3D CNN model for video classification. After that, many works extend C3D to different backbones, \eg I3D \cite{CarreiraQuo} and ResNet3D \cite{HaraCan}. In addition, some works aim to reduce the complexity of 3D CNN by decomposing the 3D convolution into 2D spatial convolution and 1D temporal convolution, \eg P3D \cite{Qiu2017Learning}, S3D \cite{DBLP:journals/corr/abs-1712-04851}, R(2+1)D \cite{DuA}.

\textbf{Action anticipation}, also aka early action prediction, aiming to predict future unseen actions with historical and current information. Many works have been developed for this tasks in recent years. For instance, Hoai \etal \cite{Hoai2012Max} propose a max-margin framework with structured SVMs to address this problem. Ryoo \etal \cite{Ryoo2012Human} develop an early action prediction system by observing some evidences from the temporal accumulated features. Yu \etal \cite{Yu2013Recognize} formulate the action prediction problem into a probabilistic framework, which aims to maximize the posterior of activity given observed frames. Aliakbarian \etal \cite{AliakbarianEncouraging} develop a multi-stage LSTM architecture that leverages context-aware and action-aware features, and introduce a novel loss function that encourages the model to predict the correct class as early as possible. Gao \etal \cite{DBLP:conf/bmvc/GaoYN17a} propose a Reinforced Encoder-Decoder (RED) network for action anticipation, which uses reinforcement learning to encourage the model to make the correct anticipations as early as possible. Ke \etal \cite{Ke_2019_CVPR} propose an attended temporal feature, which uses multi-scale temporal convolutions to process the time-conditioned observation. The widely used datasets for action anticipation, \eg, UCF-101 \cite{Soomro2012UCF101}, JHMDB-21 \cite{Jhuang2013Towards}, BIT-Interaction \cite{KongInteractive}, Sports-1M \cite{Karpathy2014Large}, include short trimmed videos, and the task mainly focuses on predicting the class of the current going action timely from only a small ratio of the observed part. Our task is different from action anticipation, we mainly focus on long and unsegmented video data, \eg TVSeries, usually with large variety of irrelevant background.

\begin{figure*}
\centering
\includegraphics[width=1.0\linewidth]{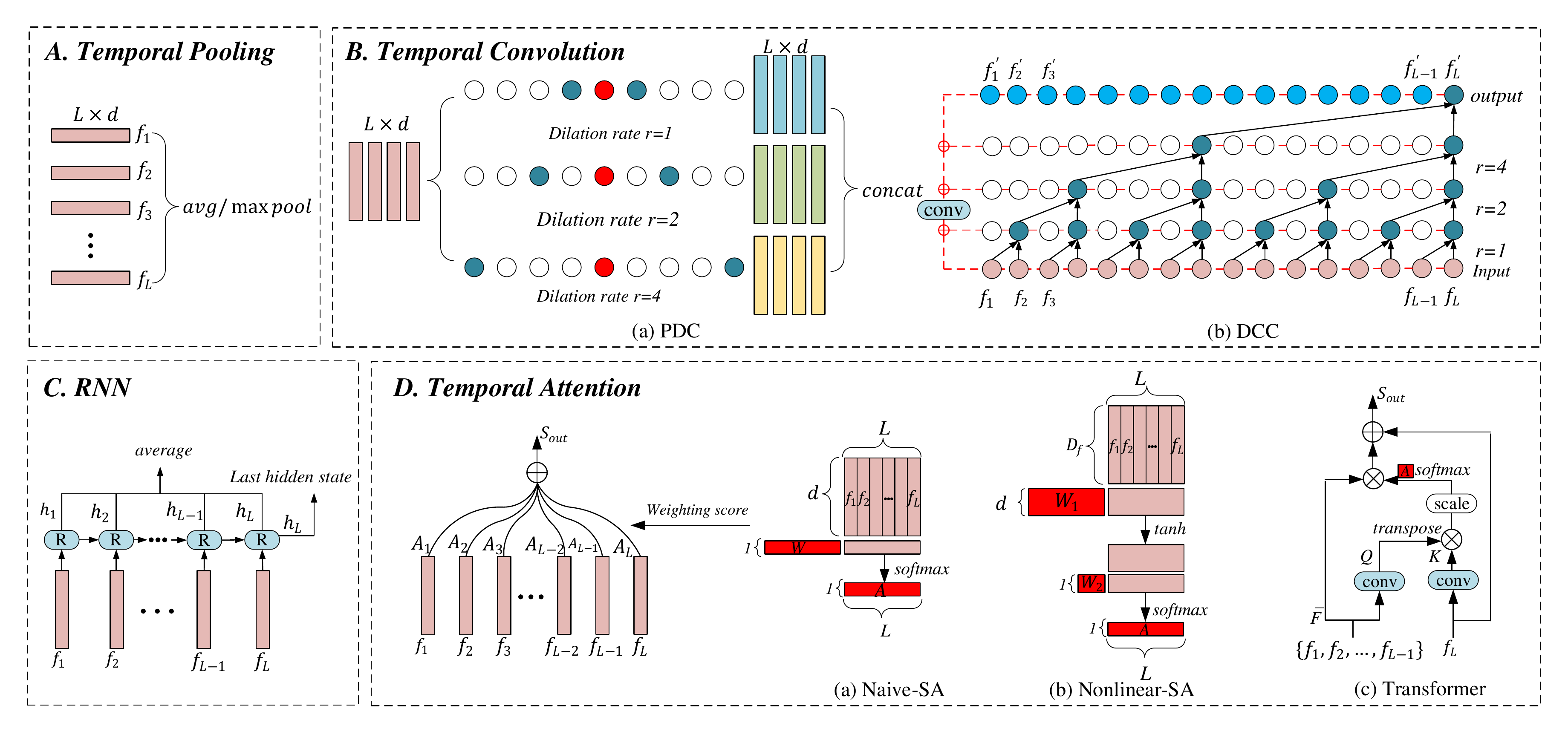}
  \caption{The temporal modeling architectures. A: Temporal pooling with max or average operation. B: Temporal convolution methods. C: Recurrent Neural Network (RNN) with LSTM or GRU cells. D: Temporal attention methods.}
\label{fig_1}
\end{figure*}

\textbf{Temporal action detection} or localization is another hot topic which aims to temporally localize and recognize actions by observing entire untrimmed videos. The main difference between this topic and OAD is the offline setting, \ie post-processing is allowed for temporal action localization. In this offline setting, the whole action can be observed first. The problem has recently received increasing attention due to its potential application in video data analysis. Shou \etal \cite{Shou2016Temporal} localize actions with three stages: action proposal generation, proposal classification and proposal regression. Xu \etal \cite{XuR} transform the Faster R-CNN \cite{Ren2015Faster} architecture into temporal action localization. Chao \etal \cite{ChaoRethinking} improve receptive field alignment using a multi-tower network and dilated temporal convolutions, and exploit the temporal context of actions for both proposal generation and action classification. Lin \etal \cite{Lin_2018_ECCV} generate proposals via learning starting and ending probability using temporal convolutional network, and achieve promising performance over previous methods. Zeng \etal \cite{Zeng_2019_ICCV} apply the Graph Convolutional Networks (GCNs) over the graph to model the relations among different proposals and learn powerful representations for the action classification and localization.

\textbf{Spatial-temporal action detection} aims to determine the precise spatial-temporal extents of actions in videos, which
has attracted increasing attention recently. Early methods mainly resort to bag-of-words representation and search spatio-temporal path. In deep learning era, many works transform image-based object detection methods into this task, \eg R-CNN~\cite{DBLP:conf/cvpr/GirshickDDM14}, Faster R-CNN \cite{Ren2015Faster}, SSD~\cite{DBLP:conf/eccv/LiuAESRFB16}, etc. These adaptive methods mainly first detect actions in frame level and then link the frame-level bounding boxes into final tubes~\cite{DBLP:journals/corr/GkioxariM14,DBLP:conf/eccv/PengS16,singh2017online,DBLP:conf/cvpr/GuSRVPLVTRSSM18}. Specially, the online setting is used in \cite{DBLP:conf/cvpr/SoomroIS16,singh2017online}.

\textbf{Online action detection} is defined as an online per-frame labelling task given streaming videos, which requires correctly classifying every frame. Geest \etal~\cite{De2016Online} first introduce the problem by introducing a realistic dataset (\ie TVSeries) and some baseline results. Their later work~\cite{De2018} introduces a two-stream feedback network, where one stream processes the input and the other one models the temporal relations. Li \etal~\cite{LiOnline} design a deep LSTM network for 3D skeletons online action detection which also estimates the start and end frame of the current action. Xu \etal~\cite{Xu_2019_ICCV} propose the Temporal Recurrent Network (TRN)  to model the temporal context by simultaneously performing online action detection and anticipation. Besides, Shou \etal~\cite{Shou2018Online} formulate the online detection of action start (ODAS) as a classification task of sliding windows and introduce a model based on Generative Adversarial Network to generate hard negative samples to improve the training of the samples.



\section{Temporal Modeling Approach}
\subsection{Problem Formulation}
Given an observed video stream $V=\{I_{0}, I_{1},\dots,I_{t}\}$ containing frames from time 0 to \textit{t}, the goal of online action detection is to recognize actions of interest occurring in frame \textit{t} with these observed frames. This is very different from other tasks like action recognition and temporal action detection which assume the entire video sequence is available at once. Formally, online action detection can be defined as the problem of maximizing the posterior probability,

\begin{equation}
    \mathbf{y}_{t}^{*}=\mathop{\arg\max}_{\mathbf{y}_{t}}P(\mathbf{y}_{t}|I_{0},I_{1},\dots,I_{t}),
\end{equation}
where $\mathbf{y}_{t}\in \mathbb{R}^{K+1}$ is the possible action label vector for frame $I_{t}$ with K action classes and one background class.  Thus, conditioned on the
observed sequence V, the action label with the maximum probability  $P(\mathbf{y}_{t}|I_{0},I_{1},...,I_{t})$ is chosen
to be the detection result of frame $I_{t}$. Generally, a pre-trained CNN model $\Phi$ is first used to extract frame-level features, \eg the feature of \textit{t}-th frame $f_t=\Phi(I_t;\theta)\in \mathbb{R}^{d}$, where $\theta$ is the fixed parameter of the model and $d$ is the dimension of feature embedding. Given the observed frame features $\{f_0,f_1,\dots, f_t\}$, a temporal modeling module aims to aggregate discriminative information from them to better estimate the output action scores.

\subsection{Temporal Modeling}
\label{sec:temporal_modeling}
For online action detection, considering that faraway frames may be unrelated to the current action state, we usually input frames of a limited sequence length $L$ to the temporal modeling module, \ie $\{f_{t-L+1},f_{t-L+2},\dots, f_t\}$. For convenience, we denote the input features as $\{f_1,f_2,\dots, f_L\}$, and assume the output of temporal modeling as $S_{out}$. Next we discuss four types of temporal modeling methods as illustrated in Fig.\ref{fig_1}.

\textbf{\emph{Temporal Pooling}}. Temporal feature pooling has been extensively used for video classification \cite{Simonyan2014Two,NgBeyond,FeichtenhoferConvolutional,KarAdaScan} which is a simple method to generate video-level representation from frame-level features. As shown in Fig.\ref{fig_1}.A, we consider two temporal pooling approaches: (1) \textit{average pooling} (AvgPool), \ie $S_{out}=\frac{1}{L}\sum _{t=1}^{L}f_{t}$, and (2) \textit{max pooling} (MaxPool) over the temporal dimension, \ie  $S_{out}=\mathop{\max_{t}}f_{t}$.

\textbf{\emph{Temporal Convolution}}. Inspired by the convolutional approaches in the analysis of temporal sequential data \cite{DauphinLanguage,BaiAn,DBLP:journals/corr/LeaFVRH16,LiGlobal,OordWaveNet} especially the WaveNet~\cite{OordWaveNet}, we evaluate (1) traditional \textit{temporal convolution} (TC), (2) \textit{pyramid dilated temporal convolution} (PDC) originally used in \cite{LiGlobal}, and (3) \textit{dilated causal convolution} (DCC) developed in \cite{OordWaveNet}. Formally, given input $F=\{f_{1},f_{2},...,f_{T}\}$, our temporal convolution models output features of the same length as follows,
\begin{equation}
\begin{matrix}
F_o=\{f^{(r)}_{o1},f^{(r)}_{o2},...,f^{(r)}_{oT}\},
\\
\\
\textrm{where}, ~~~ f_{ot}^{(r)}=\sum_{i=1}^{s}f_{[t+r\cdot i]}\times W_{[i]}^{(r)},
\end{matrix}
\end{equation}
where $r$ is a dilation rate indicating the temporal stride to sample frames, $W\in \mathbb{R}^{d\times s}$ is a convolutional kernel, and $s$ is the kernel size. It becomes our traditional temporal convolution (\ie conv1D without dilation) if $r=1$. As shown in Fig.\ref{fig_1}.B (a), PDC first separately conducts dilated temporal convolution with various dilation rates $\{r_1,r_2,\dots,r_N\}$ and then concatenates the outputs in frame-wise. Formally, the output frame-level feature $f_{t}^{'}$ of PDC is defined as follows, 
\begin{equation}
    f_{t}^{'}=concat(f_{t}^{(r_{1})},f_{t}^{(r_{2})},...,f_{t}^{(r_{N})})\in \mathbb{R}^{ Nd}.
\end{equation}
PDC uses different $r$ to cover various range of temporal context which could be better than only $r=1$. In our study, we use three dilation rates $\{1,2,4\}$ to efficiently enlarge the temporal receptive fields for PDC. As shown in Fig.\ref{fig_1}.B (b), our dilated causal convolution (DCC) stacks several dilated convolutional layers with different rates. We perform ReLU after convolution, and add a residual connection to combine the input and the output of each layer. For each layer, we increase dilation rate $r$ exponentially with the depth of the network (${\textit{i}.\textit{e}.}$, $r=O(2^{i})$ at level $i$ of the network). Specifically, we also use three dilation rates $\{1,2,4\}$ in order. To map the input sequence to an output sequence of the same length, we add zero padding with length $(s-1)*r$ in all layers. Formally, the output $f^i_t$ of the i-th layer and time $t$ is defined as,
\begin{equation}
    \begin{matrix}
    \hat{f}^{i}_t=ReLU(W^{(r_i)}f^{i-1}_t+b_{r_i}),\\
    \\
    f^{i}_t=\hat{f}^{i}_t+W_{i}f^{i-1}_t+b_{i},
    \end{matrix}
\end{equation}
where $W^{(r_i)}$ and $b_{r_i}$ are parameters for dilated convolution, $W_{i}$ and $b_{i}$ are parameters to transform $f^{i-1}_t$ for the residual connection.
After the temporal convolutional operation, we use average pooling to generate a single representation for classification by default.

\textbf{\emph{Recurrent Neural Network (RNN)}}. Recurrent Neural Network and its variants have recently been transformed from other sequence modeling topics into action classification \cite{NgBeyond,Donahue2014Long,GammulleTwo,Liu2017Global} and detection \cite{DBLP:journals/corr/abs-1903-09868,Xu_2019_ICCV,Singh2016A}. In contrast to temporal pooling operation which produces order-independent representations, RNN models the dependencies between consecutive frames and capture the temporal information of the input sequence. For each time step, the RNN cell receives the past step information $h_{t-1}$  and the current frame feature $f_{t}$, and passes the current hidden state $h_{t}$ into the next time step.

\begin{figure}
\begin{center}
\includegraphics[width=0.9\linewidth]{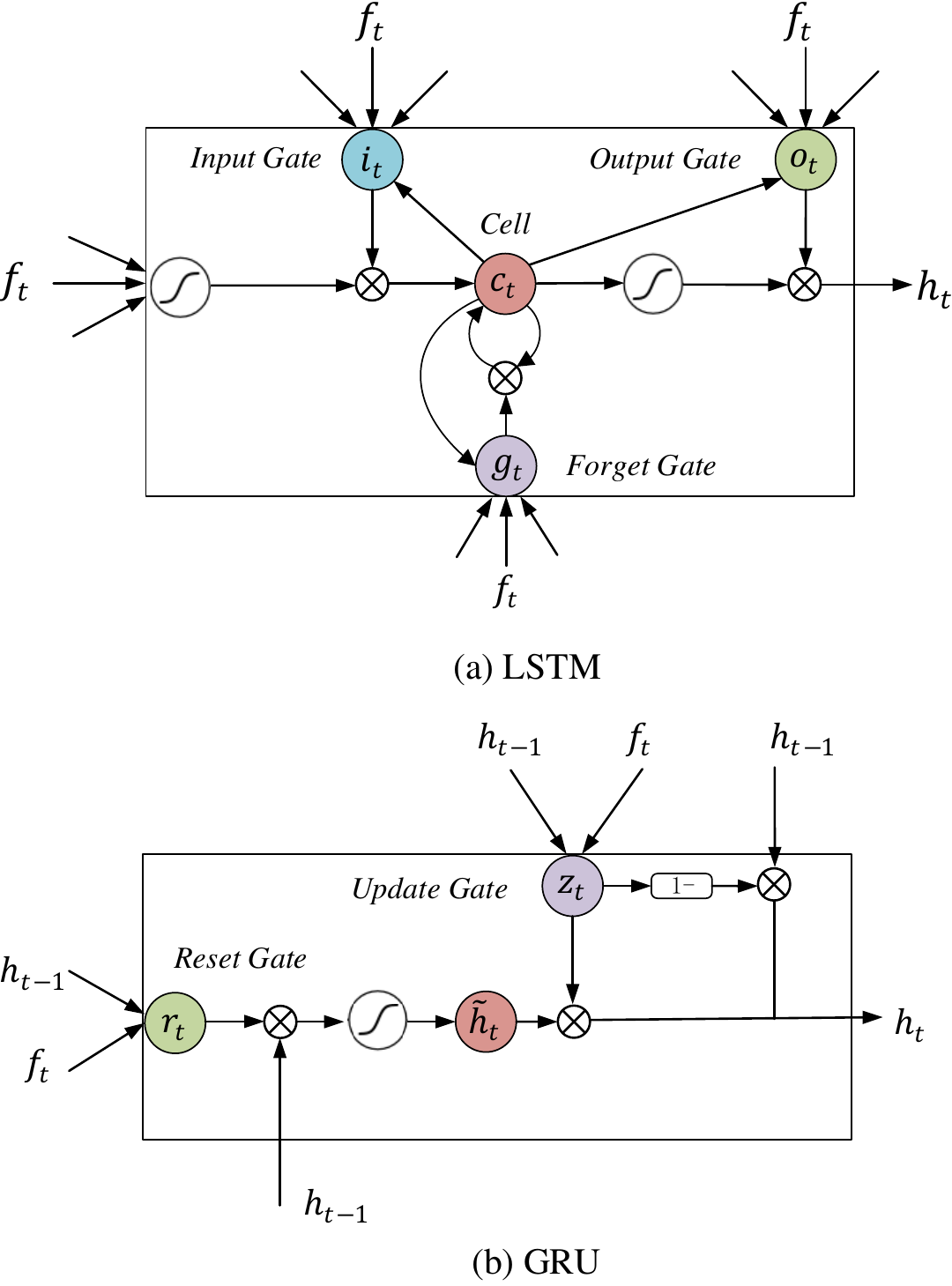}
\end{center}
  \caption{Illustration of (a) LSTM and (b) GRU. $c$, $i$, $f$, and $o$ in LSTM (a) are respectively the $memory$ $ cell$, $input~gate$, $forget~gate$ and $output~gate$. $r$ and $z$ in GRU (b) are the reset and update gates. $h$ and $\tilde {h}$ are respectively the hidden activation and the candidate activation.}
\label{fig_RNN}
\end{figure}

Specifically, we evaluate two popular recurrent cells, namely Long Short-Term Memory (LSTM) and Gated Recurrent Unit (GRU). As a special RNN structure for  sequence modeling, LSTM has been proven stable and powerful for modeling long-range dependencies in various topics \cite{Graves1997Long, DBLP:journals/corr/Graves13, DBLP:journals/corr/SutskeverVL14} and online action anticipation~\cite{DBLP:conf/bmvc/GaoYN17a}. We illustrate LSTM in Fig.\ref{fig_RNN}(a) following the implementation of \cite{DBLP:journals/corr/Graves13}. Formally, LSTM is formulated as follows,

\begin{equation}
\begin{array}{c}
i_{t}=\sigma (W_{i}f_{t}+U_{i}h_{t-1}+V_{i}c_{t-1}+b_{i}),\\
g_{t}=\sigma (W_{g}f_{t}+U_{g}h_{t-1}+V_{g}c_{t-1}+b_{g}),\\
c_{t}=g_{t}c_{t-1}+i_{t}tanh(W_{c}f_{t}+U_{c}h_{t-1}+b_{c}),\\
o_{t}=\sigma (W_{o}f_{t}+U_{o}h_{t}+V_{o}c_{t}+b_{o}),\\ h_{t}=o_{t}tanh(c_{t}),
\end{array}
\end{equation}
where $\sigma$ is the logistic sigmoid function, and $i$, $g$, $c$ and $o$ are respectively the \textit{input gate}, \textit{forget gate}, \textit{memory cell} and \textit{output gate}. $h$ is the hidden state activation vector.

 Similarly to LSTM unit, the GRU has gating units that modulate the flow of information inside the unit, as illustrated in Fig.\ref{fig_RNN}(b).  The main difference between LSTM and GRU is that there is no separate memory cell in GRN. Formally, the GRU can be formulated as follows,
 \begin{equation}
     \begin{array}{c}
r_{t}=\sigma (W_{r}f_{t}+U_{r}h_{t-1}),\\
\tilde{h}_{t}=tanh (W_{h}f_{t}+U_{h}(r_{t}\odot h_{t-1}),\\
z_{t}=\sigma (W_{z}f_{t}+U_{z}h_{t-1}),\\
h_{t}=(1-z_{t})h_{t-1}+z_{t}h_{t},
     \end{array}
 \end{equation}
where $r_{t}$ is a set of reset gates, $z_{t}$ is an update gate, and $\odot$ is an element-wise multiplication.

Since the output of RNN is another sequence, we consider two methods to generate the final single representation $S_{out}$, i) following the traditional Encoder-Decoder method, we directly take the hidden state $h_{L}$ at the last time step, \ie  $S_{out}=h_{L}$. ii) We average the outputs of all the time steps, \ie  $S_{out}=\frac{1}{L}\sum_{t=1}^{L}h_{t}$.

\textbf{\emph{Temporal Attention}}.
The attention mechanism \cite{BahdanauNeural,lin2017structured,yang2016hierarchical,DBLP:journals/corr/VaswaniSPUJGKP17} allows the model to selectively focus on only a subset of frames by increasing the attention weights of the corresponding temporal feature, while ignoring irrelevant signals and noise. We evaluate four attention methods, namely \textit{naive self-attention} (Naive-SA), \textit{nonlinear self-attention} (Nonlinear-SA) with a FC-tanh-FC-Softmax architecture, \textit{Non-local} block or standard self-attention with a skip connection, and \textit{Transformer} with the current frame as the query (Q) information. Given a feature sequence $F=\{f_{1}, f_{2},\dots,f_{L}\}$, the Naive-SA can be implemented by a linear fully-connected (FC) layer and Softmax function as follows,
\begin{equation}
\label{eq:Naive-SA}
    \bm{a}=\textrm{Softmax}(WF^{T}+b),
\end{equation}
where $W\in \mathbb{R}^d$ and $b$ are parameters of the FC, and $\bm{a}\in \mathbb{R}^L$ is the attention weight vector. Similar to \cite{yang2016hierarchical}, we can also add more nonlinear operation as follows (\ie the Nonlinear-SA),
\begin{equation}
\label{eq:Nonlinear-SA}
    \bm{a}=\textrm{Softmax}(U_{2}tanh(U_{1}F^{T}+\mathbf{b}_{1})+b_{2}),
\end{equation}
where $U_{1}\in \mathbb{R}^{d_1 \times d}$ is a weight matrix, $\mathbf{b}_{1}\in \mathbb{R}^{d_1}$ is the bias vector, and $U_{2}\in \mathbb{R}^{d_1}$ and $b_{2}$ are parameters of the second FC layer. With the attention weights, the output representation is the weighted average vector $S_{out}=\sum_{t=1}^{L} \bm{a}_t f_{t}$.

\textit{Transformer} is another popular attention based model, which was originally proposed to replace traditional recurrent models for machine translation~\cite{DBLP:journals/corr/VaswaniSPUJGKP17}. The core idea of Transformer is to model correlation between contextual signals by an attention mechanism. Specifically, it aims to encode the input sequence to a higher-level representation by modeling the relationship between queries ($Q$) and memory (keys ($K$) and values ($V$)) with,
\begin{equation}
\label{lable:transformer}
    A =Attention(Q, K, V)=\textrm{Softmax}(\frac{QK^{T}}{\sqrt{d_{m}}})V,
\end{equation}
where $Q\in R^{L_q\times d_{m}}$, $K\in R^{L_k\times d_{m}}$ and $V\in R^{L_k\times d_{v}}$. This architecture becomes standard ``self-attention'' with $Q=K=V=\{f_1,f_2,\cdots, f_L\}$. Normally, we use two convolution layers followed by Batch Normalization and ReLU to generate two new features $Q$ and $K$ from $F$, and the \textit{Non-local} method~\cite{Wang2017Non} further adds a skip connection between the input and the output as follows,
\begin{equation}
\label{eq:Non-local}
    F^{'}=AF+F, F^{'}\in \mathbb{R}^{L\times d}.
\end{equation}
The updated temporal feature $F^{'}$ is processed with average pooling by default to generate the final temporal representation $S_{out}=Avg(F')\in \mathbb{R}^{d}$.

The query $Q$ in Eq. (\ref{lable:transformer}) can be also a single feature vector, similar to \cite{Wu_2019_CVPR} which replaces the self-attention weights by the one between local feature and long-term features, we compute the dot-product attention between current feature $f_{L}$ and historical features $\bar{F}=\{f_{1},f_{2},...,f_{L-1}\}$ as illustrated in Fig.\ref{fig_1}.D (c). This adaption is based on the assumption that the current frame is the most important one for online action detection. With this operation, an attention weight vector $\bm{a}\in \mathbb{R}^{L-1}$ is obtained and used to get the final representation as,
\begin{equation}
\label{eq:transformer}
    S_{out}=\bm{a} \bar{F}+f_t, S_{out}\in \mathbb{R}^{d}.
\end{equation}

\textbf{\emph{Training and Inference}}.
With the output $S_{out}$ of temporal modeling module, we use linear FC layer with Softmax for classification, and train the whole network with cross-entropy loss. Specifically, we divide the feature sequence of a video into non-overlapped windows (size $L$) as the input of our temporal modeling module. At test stage, sliding window (size $L$) with stride 1 is used to formulate the input, and the prediction is made for the last frame.

\section{Experimental Configuration}
In this section, we first introduce two widely-used OAD datasets, \ie TVSeries and THUMOS-14, and then describe our implementation details, including unit-level feature extraction and hyperparameter settings.
\subsection{Datasets}
\textbf{TVSeries} \cite{De2016Online} is originally proposed for online action detection, which consists of $27$ episodes of $6$ popular TV series, namely \textit{Breaking Bad} ($3$ episodes), \textit{How I Met Your Mother} ($8$), \textit{Mad Men} ($3$), \textit{Modern Family} ($6$), \textit{Sons of Anarchy} ($3$), and \textit{Twenty-four} ($4$). It contains totally $16$ hours of video. The dataset is temporally annotated at the frame level with $30$ realistic, everyday actions (${\textit{e}.\textit{g}.}$, \textit{pick up}, \textit{open door}, \textit{drink}, etc.). It is challenging with diverse actions, multiple actors, unconstrained viewpoints, heavy occlusions, and a large proportion of non-action frames.

\begin{figure}
\begin{center}
\includegraphics[width=0.9\linewidth]{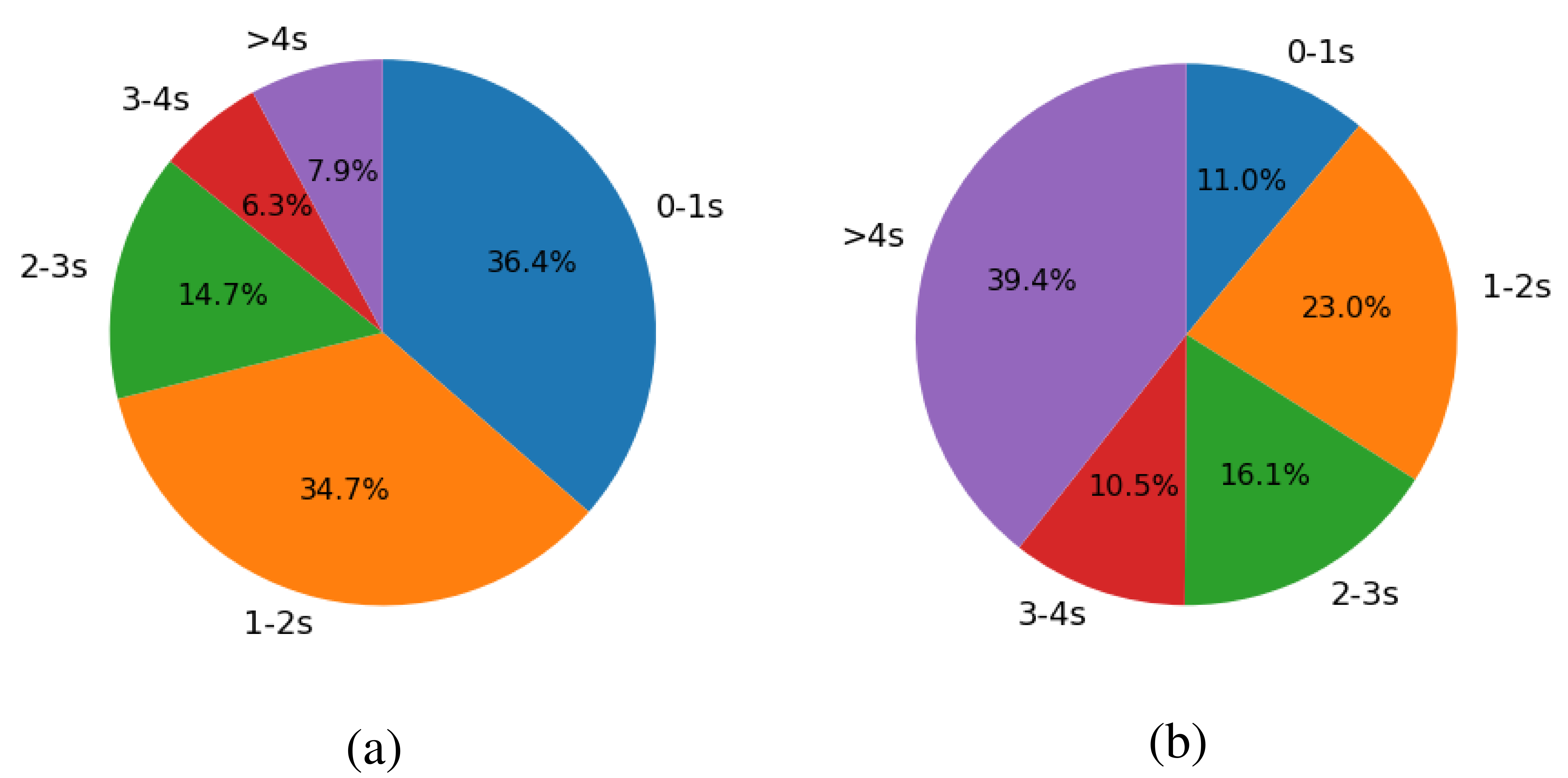}
\end{center}
  \caption{The temporal length distributions of action instances on (a) TVSeries and (b) THUMOS-14.}
\label{fig_dataset}
\end{figure}

\textbf{THUMOS-14} \cite{THUMOS14} is a popular benchmark for temporal action detection. It contains over $20$ hours of sport videos annotated with $20$ actions. The training set (${\textit{i}.\textit{e}.}$ UCF101 \cite{Soomro2012UCF101}) contains only trimmed videos that cannot be used to train temporal action detection models. Following prior works \cite{DBLP:conf/bmvc/GaoYN17a, Xu_2019_ICCV}, we train our model on the validation set (including $3K$ action instances in $200$ untrimmed videos) and evaluate on the test set (including $3.3K$ action instances in $213$ untrimmed videos).

To investigate the characters of the used datasets, we depict the temporal length distributions of action instances on TVSeries and THUMOS-14 in Fig.\ref{fig_dataset}. We observe that 70\% of action instances are very short on TVSeries (\ie 0-2s) while half of instances are longer than 3 seconds on THUMOS-14.

\subsection{Evaluation Protocols}
For each class on TVSeries, we use the per-frame calibrated average precision (cAP) which is proposed in \cite{De2016Online},
\begin{equation}
    cAP=\frac{\sum \nolimits_{k}cPrec(k)\ast I(k)}{P},
\end{equation}
where calibrated precision $cPrec=\frac{TP}{TP+FP/w}$, $I(k)$ is an indicator function that is equal to $1$ if the cut-off frame $k$ is a true positive, $P$ denotes the total number of true positives, and $w$ is the ratio between negative and positive frames. The mean cAP over all classes is reported for final performance. The advantage of cAP is that it is fair for class imbalance condition. For THUMOS-14, we report per-frame mean Average Precision (mAP) performance.

\subsection{Implementation details}
\label{sec_4.3}
\textbf{Unit-level feature extraction.}
Following previous work \cite{Xu_2019_ICCV,DBLP:conf/bmvc/GaoYN17a,Gao_2018_ECCV,Lin_2018_ECCV}, a long untrimmed video is first cut into video units without overlap, each unit contains $n_{u}$ continuous frames. A video chunk $u$ is processed by a visual encoder $E_{v}$ to extract the unit-level representation $f_{u}=E_{v}(u)\in \mathbb{R}^{d}$. In our experiments, we extract frames from all videos at $24$ frames per second. The video unit size $n_{u}$ is set to $6$, ${\textit{i}.\textit{e}.}$ $0.25$ second. We use two-stream \cite{Xiong2016CUHK} network as the visual encoder $E_{v}$ that is pre-trained on ActivityNet-1.3 \cite{Caba2015ActivityNet}. In each unit, the central frame is sampled to calculate the appearance CNN feature, it is the \textit{Flatten 673 layer} of ResNet-200 \cite{DBLP:journals/corr/HeZRS15}. For the motion feature, we sample $6$ consecutive frames at the center of a unit and calculate optical flows between them. These flows are then fed into the pretrained BN-Inception model \cite{Ioffe2015Batch}, and the output of \textit{global pool layer} is extracted. The motion features and the appearance features are both 2048-D, and are concatenated into 4096-D vectors (\ie $d= 4096$), which are used as unit-level features.

\textbf{Hyperparameter setting.}
For the PDC model, the concatenate features are fed into an addition $1\times 1$ convolution to reduce the feature dimensions to $4096$. For the DCC model, we use 3 dilated convolution layers, each of which is comprised of one dilated convolution with kernel size $s=2$, followed by a ReLU and dropout. The output dimension of the second layer is set to $2048$, and thus a $1\times 1$ convolution is added for residual connection. Our experiments are conducted in Pytorch. We use SGD optimizer to train the network from scratch. The leaning rate, momentum and decay rate are set to $10^{-3}$, $0.9$ and $0.95$, respectively. All of our experiments are implemented with 8 GTX TITAN X GPU, Intel i7 CPU, and 128GB memory.

\section{Exploration of Temporal Modeling Methods}
In this section, we first present a quick comparison among the best settings of the four mentioned temporal modeling methods, and then extensively explore both individual temporal modeling methods and their combinations, and finally compare our results to the state of the arts.

\subsection{A Quick Comparison of Temporal Modeling Methods}
As mentioned in the Introduction, we totally explore eleven temporal modeling methods from four meta types, namely \textit{temporal pooling}, \textit{temporal attention}, \textit{RNN}, and \textit{temporal convolution}. For a quick glance, Table \ref{tab:quick} presents the results of the best choice (\ie the 2nd row) for each meta type. For a fair comparison, the input sequence length L is fixed as 4.
Several observations can be concluded as following. First, temporal convolution (\ie DCC) achieves the best results on both TVSeries and THUMOS-14, which indicates that discriminative information can be obtained effectively by temporal convolution. Second, temporal attention (\ie Transformer) performs slightly better than temporal pooling (\ie AvgPool), which demonstrates the effectiveness of attention mechanism. Third, RNN (\ie LSTM) outperforms Transformer and AvgPool with sizable margins on both datasets, which shows that the temporal dependencies captured by LSTM is crucial for accurate online action detection. Overall, an interesting finding is that the temporal-dependent methods, \ie temporal convolution and RNN, are superior to these temporal-independent methods for online action detection.

\begin{table}[t]
\centering
\small
\caption{A quick comparison among the best settings of different meta types of temporal modeling methods. The best choice of each type is presented in the 2nd row. }
\begin{tabular}{ccccc}
\toprule
 & Pooling & Attention & RNN & Convolution\\ \midrule
  & AvgPool & Transformer & LSTM & DCC \\ \midrule
TVSeries & 81.2 & 81.5 & 82.9 & 83.1 \\
THUMOS-14 & 41.5 & 43.3 & 45.9 & 46.8 \\ \bottomrule
\end{tabular}
\label{tab:quick}
\end{table}

\subsection{Ablation Study for Individual Temporal Modeling Method}
\begin{figure}[t]
\begin{center}
\includegraphics[width=1.0\linewidth]{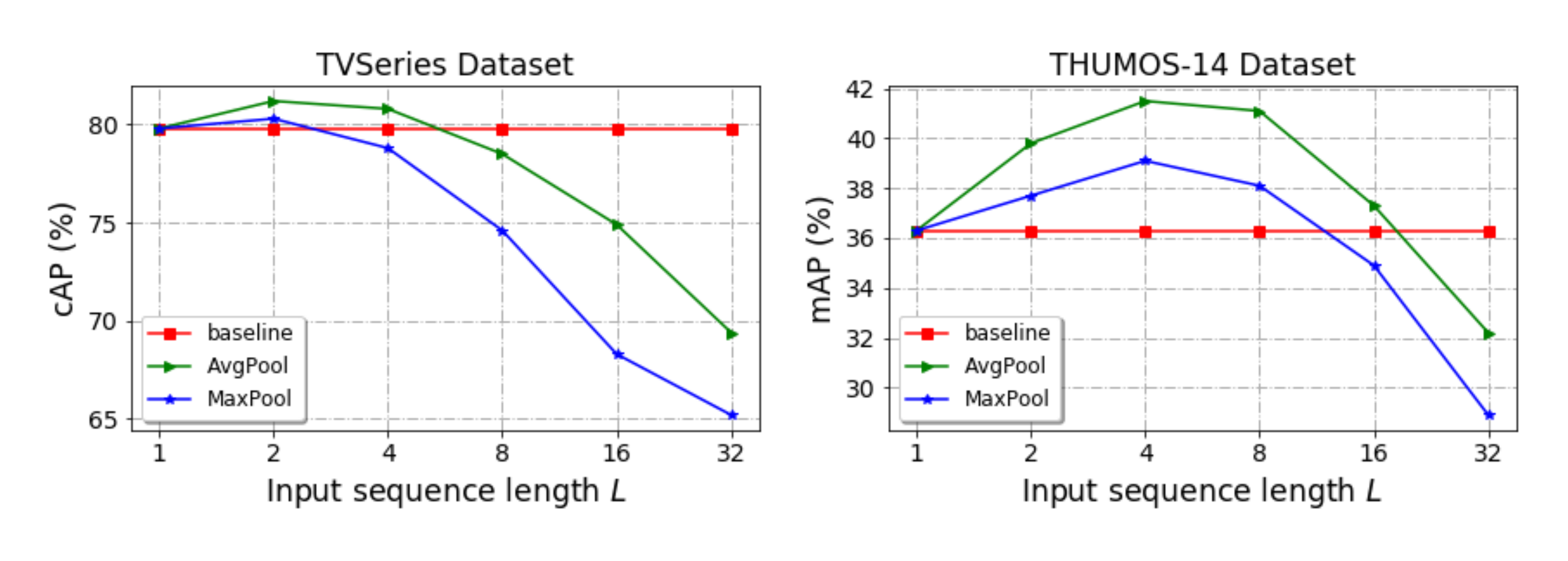}
\end{center}
  \caption{Comparison between average pooling and max pooling with different sequence length $L$ as input.}
\label{fig:pooling}
\end{figure}

\begin{figure}[t]
\begin{center}
\includegraphics[width=1.0\linewidth]{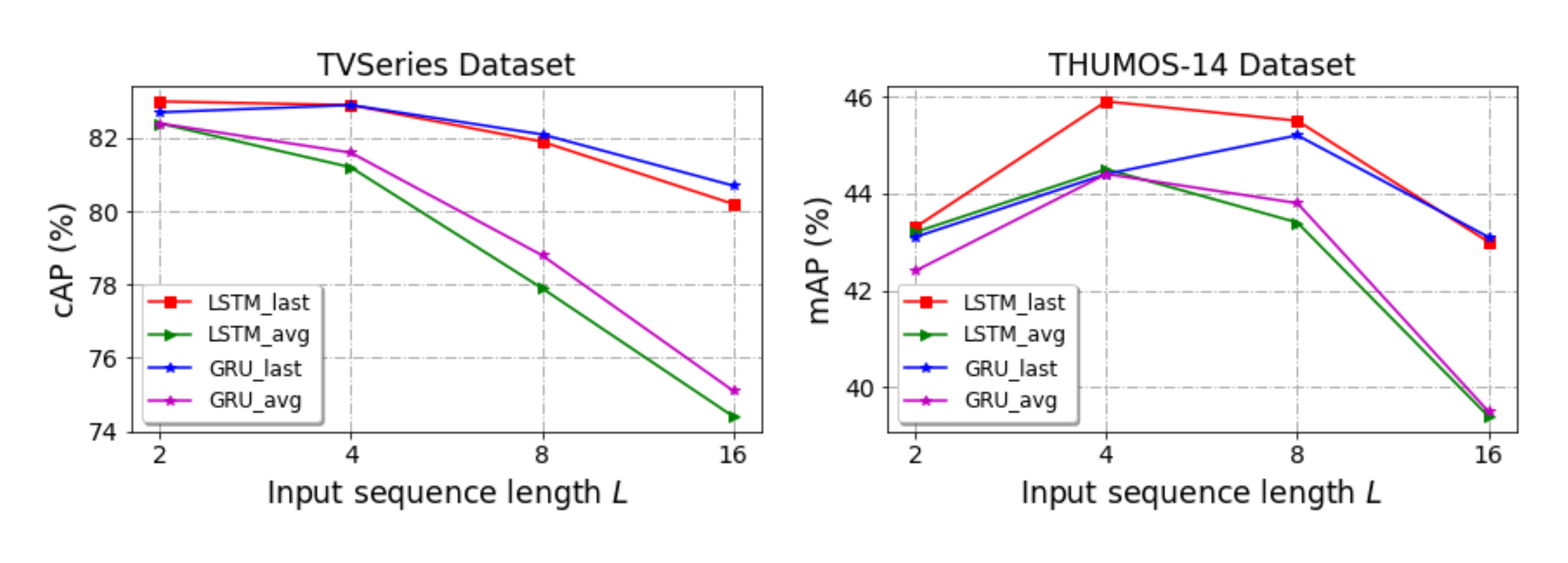}
\end{center}
  \caption{Evaluation of different sequence length $L$ with different output strategies for LSTM and GRU.}
\label{fig:output_select}
\end{figure}

\textbf{Temporal pooling}.
We test two temporal pooling methods (\ie average pooling and max pooling) with different sequence lengths. The results are shown in Fig.\ref{fig:pooling}. We also compare them to the baseline that uses a fully-connected (FC) layer and Softmax to generate action probabilities frame by frame. The baseline model only relies on the current frame feature which obtains 79.8\% (cAP) and 36.3\% (mAP) on TVSeries and THUMOS-14, respectively. For temporal pooling, it is clear that average pooling consistently performs better than max pooling on both datasets. Increasing the sequence length improves both pooling methods in the beginning and degrades them dramatically after the saturation length. This can be explained by that appropriate historical information introduces useful context for online action detection while long-term historical information may introduce unrelated information and may also smooth the final representation.
Another observation is that increasing the sequence length after L=4 is seriously harmful for TVSeries while not for THUMOS-14. This effect indicates a fact that each video in TVSeries contains multiple actions and numerous varied background frames while each video in THUMOS-14 only contains one action instance. Overall, the simple AvgPool method (L=4) respectively improves the baselines on TVSeries and THUMOS-14 by 1.4\% and 5.2\%.

\textbf{RNN}. We evaluate LSTM and GRU in the following four aspects: input sequence length, output strategy, hidden size, and the number of recurrent layers.

\textit{Input sequence length and output strategy.} For these two factors, we vary the sequence length from 2 to 16, and evaluate two alternative output strategies including the last hidden state $S_{out}=h_{L}$ and the average hidden state $S_{out}=\frac{1}{L}\sum_{t=1}^{L}h_{t}$. The hidden size is fixed to 4096 and only one recurrent layer is used for this evaluation. Fig.\ref{fig:output_select} illustrates the comparison results for LSTM and GRU.
Several conclusions can be drawn as following. First, the `last hidden state' strategy performs consistently better than the `average hidden state'. It can be explained by that both LSTM and GRU automatically accumulate discriminative information into the last state by their temporal dependency operations while averaging all the hidden states may introduce unrelated or noisy information for online action detection. Second, LSTM performs better than GRU on THUMOS-14 while similarly or worse on TVSeries. This indicates that the separate memory cell in LSTM is helpful to capture more context information which is crucial for THUMOS-14 while too much context (unrelated actions or background) can degrade performance on TVSeries. Third, the effect of sequence length for both LSTM and GRU is the same as the one for pooling methods, and the best trade-off sequence length is 4 on both datasets.

\begin{figure}[t]
\begin{center}
\includegraphics[width=1.0\linewidth]{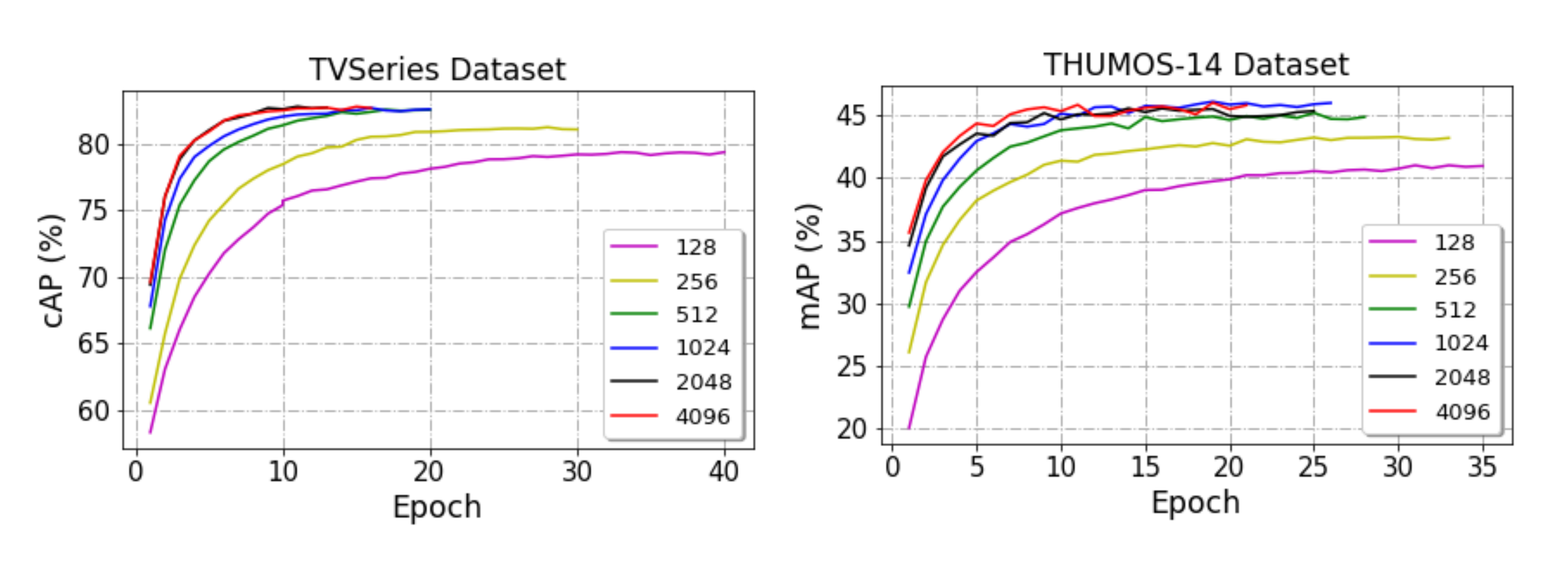}
\end{center}
  \caption{Evaluation of hidden size for LSTM with the last hidden state output strategy and sequence length $L=4$.}
\label{fig:hidden_size}
\end{figure}

\textit{Hidden size.} We choose LSTM, and test different hidden sizes $D_{h}=128$, $256$, $512$, $1024$, $2048$, $4096$.  The last hidden state output strategy and sequence length $L=4$ are used for this evaluation. The  results are shown in Fig.\ref{fig:hidden_size}. A clear observation is that increasing hidden size improves the final performance significantly on both datasets. In addition, it gets saturated after hidden size 2048, and the best results are respectively 82.9\% and 45.9\% on TVSeries and THUMOS-14 with hidden size 4096.

\begin{figure}[t]
\begin{center}
\includegraphics[width=1.0\linewidth]{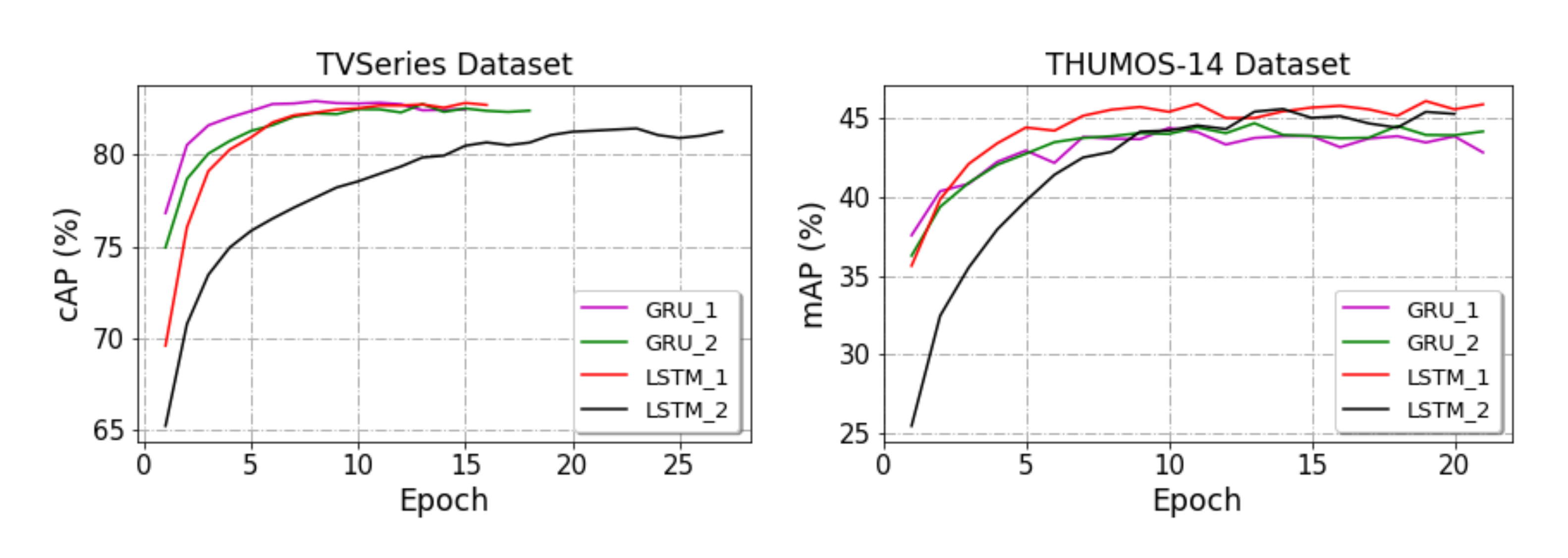}
\end{center}
  \caption{Evaluation of different number of layers for LSTM and GRU with the last hidden state, sequence length $L=4$, and hidden size $D_{h}=4096$.}
\label{fig_3}
\end{figure}



\textit{The number of recurrent layers.} Generally, one can easily stack several recurrent layers to model the complex dependency of sequences. To this end, we evaluate the number of recurrent layers for both LSTM and GRU on TVSeries and THUMOS-14. The results are shown in Fig.\ref{fig_3}.
Interestingly, adding one more layer does not bring performance gain and even dramatically degrades the performance for LSTM on both datasets. The main problem is that adding one more recurrent layer can double the number of parameters leading to overfit easily.

\textbf{Temporal convolution}.
As shown in Table \ref{tab_4}, we compare temporal convolution models with different kernel size $s$ and dilation rate $r$, denoted as $(s, r)$. For PDC and DCC, we use temporal convolutional filters with kernel size $s=2$ as a building block. The input sequence length is fixed to $L=4$ for all the comparison experiments. In order to obtain output with equal length as the input, we add zero padding as it needs. Several observations can be concluded as follows. First, the comparison between TC(2,1) and TC(3,1) indicates that the kernel size $s=2$ is slightly better than $s=3$ on both datasets. Second, the comparison among TC(2,1), TC(2,2), and TC(2,4) shows that different dilation rates perform similarly on both datasets. Third, both PDC and DCC which combines TC(2,1), TC(2,2), and TC(2,4) in either parallel or serial manner significantly improve the traditional TC models, and DCC performs best. This demonstrates that combining multi-dilation temporal convolution layers can capture complementary multi-scale action information.

\begin{table}[t]
\footnotesize
\centering
\caption{Comparison between different temporal convolutional models.}
\label{tab_4}
\resizebox{\linewidth}{!}{
\begin{tabular}{ccccccc}
\toprule
 & TC(2,1) & TC(3,1) & TC(2,2) & TC(2,4) & PDC & DCC \\ \midrule
TVSeries & 81.1 & 80.9 & 81.0 & 80.8 & 82.7 & \bf{83.1} \\
THUMOS-14 &  42.4 & 41.9 & 42.6 & 42.7 & 46.1 & \bf{46.8} \\ \bottomrule
\end{tabular}}
\end{table}

\begin{table}[]
\footnotesize
\centering
\caption{Comparison between different temporal attention models.}
\begin{tabular}{ccccc}
\toprule
 & Naive-SA & Nonlinear-SA & Non-local & Transformer \\ \midrule
TVSeries & 80.1 & 80.9 & 80.9 & 81.5 \\
THUMOS-14 & 39.9 & 42.5 & 42.4 & 43.3 \\ \bottomrule
\end{tabular}
\label{tab:attention_models}
\end{table}

\begin{table}[t]
\tiny
\centering
\caption{Evaluation of $d_{1}$ on TVSeries (cAP \%) and THUMOS-14 dataset (mAP \%).}
\resizebox{\linewidth}{!}{
\begin{tabular}{cccccc}
\toprule
$d_{1}$ & 256 & 512 & 1024 & 2048 & 4096 \\ \midrule
TVSeries & 80.6 & 80.9 & 80.6 & 80.7 & 80.3 \\
THUMOS-14 & 42.0 & 42.0 & 42.5 & 41.1 & 41.5 \\ \bottomrule
\end{tabular}
\label{tab:d1_Nonlinear}}
\end{table}

\textbf{Temporal attention}.
We compare four different attention models mentioned in Sec.\ref{sec:temporal_modeling}, \ie \textit{Naive-SA} as described in Eq.(\ref{eq:Naive-SA}), \textit{Nonlinear-SA} as described in Eq.(\ref{eq:Nonlinear-SA}), \textit{Non-local} as described in Eq.(\ref{eq:Non-local}), and \textit{Transformer} as described in Eq.(\ref{eq:transformer}). As shown in Table \ref{tab:attention_models}, several observations can be concluded as following. First, Nonlinear-SA outperforms Naive-SA by $0.8\%$ on TVSeries and $2.6\%$ on THUMOS-14. Compared to Naive-SA, Nonlinear-SA computes attention weights with one more nonlinear \textit{tanh} and linear FC which may be more effective for modeling the complex temporal relationships.
Second, Non-local performs equally to Nonlinear-SA on both datasets, indicating that they share the similar attention mechanism more or less. Third, Transformer with current frame feature as a query performs better than Non-local by $0.6\%$ on TVSeries and $0.9\%$ on THUMOS-14, showing the effectiveness of our proposed design (\ie computing the attention between current frame feature with historical features) for online action detection.

As there is a hyper parameter $d_{1}$ in Nonlinear-SA (see Eq.(\ref{eq:Nonlinear-SA})) which can impact the final performance, we also conduct an evaluation in Table \ref{tab:d1_Nonlinear}. We observe $d_{1}=512$ (1024) yield the best performance for TVSeries (THUMOS-14), and the final performance is not very sensitive to it. 

\begin{figure*}
\begin{center}
\includegraphics[width=1.0\linewidth]{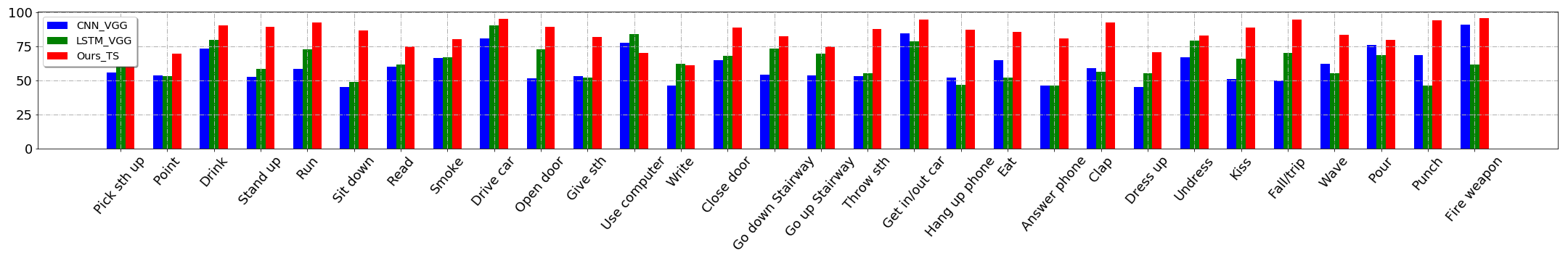}
\end{center}
  \caption{The online detection results of ours compared to previous methods in terms of per-frame cAP ($\%$) for each action class on TVSeries.}
\label{fig:per_class}
\end{figure*}

\begin{table}[]
\footnotesize
\caption{Comparison between different hybrid temporal models on TVSeries (cAP \%) and THUMOS-14 (mAP \%).}
\label{tab:hybrid_models}
\resizebox{\linewidth}{!}{
\begin{tabular}{c|lcc}
\toprule
 &Hybrid models & TVSeries & THUMOS-14 \\ \midrule
M1 & LSTM $\oplus$ Transformer & 83.6 & 47.7 \\
M2 & DCC $\oplus$ Transformer & \textbf{84.3} & 47.1 \\
M3 & LSTM $\oplus$ DCC $\oplus$ Transformer & 83.0 & 48.5 \\
M4 & DCC $\oplus$ LSTM $\oplus$ Transformer & 83.7 & \textbf{48.6} \\
M5 & DCC $\oplus$ LSTM & 83.2 & 47.9\\
M6 & LSTM $\oplus$ DCC $\oplus$ AvgPool & 81.5 & 47.5 \\ \bottomrule
\end{tabular}}
\end{table}

\begin{table}[t]
\centering
\caption{Comparison with state-of-the-art methods in terms of per-frame cAP (\%) on TVSeries.}
\small
\resizebox{\linewidth}{!}{
\begin{tabular}{ccc}
\toprule
Method & Inputs & cAP \\ \midrule
CNN (De Geest \etal, 2016) \cite{De2016Online} & \multirow{6}{*}{VGG} & 60.8 \\
LSTM (De Geest \etal, 2016) \cite{De2016Online}&  & 64.1 \\
RED (Gao \etal, 2017) \cite{DBLP:conf/bmvc/GaoYN17a}&  & 71.2 \\
Stacked LSTM (De Geest and Tuytelaars, 2018) \cite{De2018}&  & 71.4 \\
2S-FN (De Geest and Tuytelaars, 2018) \cite{De2018} &  & 72.4 \\
TRN (Xu \etal, 2019) \cite{Xu_2019_ICCV} &  & 75.4 \\ \hline
SVM (e Geest \etal, 2016) \cite{De2016Online} & FV & 74.3 \\ \hline
RED (Gao \etal, 2017) \cite{DBLP:conf/bmvc/GaoYN17a} & \multirow{2}{*}{TS} & 79.2 \\
TRN (Xu \etal, 2019) \cite{Xu_2019_ICCV} &  & 83.7 \\ \hline
Ours  & TS & \bf{84.3} \\ \bottomrule
\end{tabular}}
\label{tab:tvseries_soa}
\end{table}

\begin{table}[t]
\centering
\caption{Comparison with published state-of-the-art methods in terms of per-frame mAP (\%) on THUMOS-14.}
\tiny
\resizebox{\linewidth}{!}{
\begin{tabular}{cc}
\toprule
Method & mAP \\ \midrule
Single-frame CNN (Simonyan and Zisserman, 2014) \cite{Simonyan2014Very}& 34.7 \\
Two-stream CNN (Simonyan and Zisserman, 2014) \cite{Simonyan2014Two} & 36.2 \\
C3D+LinearInterp (Shou \etal, 2017) \cite{ShouCDC}& 37.0 \\
Predictive-corrective (Dave \etal, 2017) \cite{DavePredictive} & 38.9 \\
LSTM (Donahue \etal, 2014) \cite{Donahue2014Long} &  39.3 \\
MultiLSTM (Yeung \etal, 2015) \cite{Yeung2015Every} & 41.3 \\
CDC (Shou \etal, 2017) \cite{ShouCDC}& 44.4 \\
RED (Gao \etal, 2017)\cite{DBLP:conf/bmvc/GaoYN17a} & 45.3 \\
TRN (Xu \etal, 2019) \cite{Xu_2019_ICCV}&  47.2 \\ \hline
Ours & \bf{48.6} \\ \bottomrule
\end{tabular}}
\label{tab:thumos_soa}
\end{table}

\subsection{Combination of Temporal Modeling Methods}
Generally, these sequence-to-sequence temporal models, \eg DCC and LSTM can be further processed by aggregation methods like temporal pooling and temporal attention to generate a single representation. Thus, we present several hybrid temporal modeling methods which combine different temporal modeling methods, aiming to uncover the complementarity among them. Specifically, according to their characters, we mainly combine those temporal-dependent models and temporal-independent models as follows.
\begin{itemize}
\item[M1] LSTM $\oplus$ Transformer: The hidden states at all time steps are further fed into Transformer to generate a single representation, and the classification is performed on the representation.
\item[M2] DCC $\oplus$ Transformer: The output of the DCC network is the same as the input sequence length, and Transformer is performed on the output sequence to generate a single representation for classification.
\item[M3] DCC $\oplus$ LSTM $\oplus$ Transformer: The output sequence is further processed by LSTM aiming to capture strong temporal dependency, and finally Transformer is used to generate the representation for classification.
    \item[M4] LSTM $\oplus$ DCC $\oplus$ Transformer: The hidden states are first fed into DCC and then Transfomer. This model is similar to M3 except for that it swaps the order of DCC and LSTM;

    \item[M5] DCC $\oplus$ LSTM: The output sequence of DCC is processed by LSTM, and then the last hidden state is used for action classification. 
    \item[M6] LSTM $\oplus$ DCC $\oplus$ AvgPool: This model replaces the Transformer of M3 with AvgPool to generate the final representation for classification. 
\end{itemize}

The results of the hybrid methods on TVSeries and THUMOS-14 are shown in Table \ref{tab:hybrid_models}. Several observations can be concluded as following. First, the best results on TVSeries and THUMOS-14 are achieved by M2 and M4, respectively. Second, combining temporal-dependent models (\ie LSTM and DCC) with temporal-independent ones (\ie Transformer) largely improves individual models, which indicates that they are complementary. Third, integrating LSTM into DCC $\oplus$ Transformer (\ie M2$\rightarrow$M3) degrades the performance by 1.3\% on TVSeries while increases the one by 1.4\% on THUMOS-14. This may be explained by that temporal dependencies are important for these long-term action instances of THUMOS-14 while harmful for the dominated short-term action instances of TVSeries.
%

\subsection{Comparison with state-of-the-art}
We compare our best results to the state-of-the-art approaches on TVSeries and THUMOS-14 in Table \ref{tab:tvseries_soa} and Table \ref{tab:thumos_soa}, respectively. With two-stream features, we achieve $84.3\%$ in terms of mean cAP on TVSeries and 48.6\% mAP on THUMOS-14, which outperforms the recent sophisticated-designed TRN \cite{Xu_2019_ICCV} by $0.6\%$ and $1.4\%$, respectively. Besides, we also present the comparison of ours with previous methods \cite{De2016Online} for each action class on TVSeries in Fig.\ref{fig:per_class}. Our method can always outperform CNN and LSTM by a large margin except for action class \textit{Use computer} and \textit{Write}.


\section{Conclusions}
In this paper, we provide a comprehensive study on temporal modeling for online action detection including four meta types of temporal modeling methods, \ie temporal pooling, temporal convolution, recurrent neural networks, and temporal attention. We extensively explore eleven individual temporal modeling methods and explore several hybrid temporal models which combine different temporal modeling methods to uncover the complementarity among them. Based on our comprehensive study, we find that a simple fusion between dilated causal convolution and Transformer or LSTM improves the individual models significantly and also outperforms the best existing performance with a sizable margin on both TVSeries and THUMOS-14 datasets.

\bibliography{refs}

\begin{thebibliography}{100}

\bibitem{AliakbarianEncouraging}
Mohammad~Sadegh Aliakbarian, Fatemeh~Sadat Saleh, Mathieu Salzmann, Basura
  Fernando, Lars Petersson, and Lars Andersson.
\newblock Encouraging lstms to anticipate actions very early.
\newblock In {\em ICCV}, 2017.

\bibitem{Aliakbarian2017Encouraging}
Mohammad~Sadegh Aliakbarian, Fatemehsadat Saleh, Mathieu Salzmann, Basura
  Fernando, and Lars Andersson.
\newblock Encouraging lstms to anticipate actions very early.
\newblock In {\em ICCV}, 2017.

\bibitem{Ba2016Layer}
Jimmy~Lei Ba, Jamie~Ryan Kiros, and Geoffrey~E. Hinton.
\newblock Layer normalization.
\newblock 2016.

\bibitem{DBLP:journals/corr/BaekKK16}
Seungryul Baek, Kwang~In Kim, and Tae{-}Kyun Kim.
\newblock Real-time online action detection forests using spatio-temporal
  contexts.
\newblock {\em CoRR}, 2016.

\bibitem{BahdanauNeural}
Dzmitry Bahdanau, Kyunghyun Cho, and Yoshua Bengio.
\newblock Neural machine translation by jointly learning to align and
  translate.
\newblock 2014.

\bibitem{BaiAn}
Shaojie Bai, J.~Zico Kolter, and Vladlen Koltun.
\newblock An empirical evaluation of generic convolutional and recurrent
  networks for sequence modeling.
\newblock 2018.

\bibitem{Caba2015ActivityNet}
Fabian Caba, Victor Escorcia, Bernard Ghanem, and Juan~Carlos Niebles.
\newblock Activitynet: A large-scale video benchmark for human activity
  understanding.
\newblock In {\em CVPR}, 2015.

\bibitem{CarreiraQuo}
Joao Carreira and Andrew Zisserman.
\newblock Quo vadis, action recognition? a new model and the kinetics dataset.
\newblock In {\em CVPR}, 2017.

\bibitem{Carreira_2017_CVPR}
Joao Carreira and Andrew Zisserman.
\newblock Quo vadis, action recognition? a new model and the kinetics dataset.
\newblock In {\em CVPR}, 2017.

\bibitem{ChaoRethinking}
Yu~Wei Chao, Sudheendra Vijayanarasimhan, Bryan Seybold, David~A. Ross, Jia
  Deng, and Rahul Sukthankar.
\newblock Rethinking the faster r-cnn architecture for temporal action
  localization.
\newblock In {\em CVPR}, 2018.

\bibitem{DBLP:journals/corr/ChoMBB14}
KyungHyun Cho, Bart van Merrienboer, Dzmitry Bahdanau, and Yoshua Bengio.
\newblock On the properties of neural machine translation: Encoder-decoder
  approaches.
\newblock {\em CoRR}, abs/1409.1259, 2014.

\bibitem{DBLP:journals/corr/ChungGCB14}
Junyoung Chung, {\c{C}}aglar G{\"{u}}l{\c{c}}ehre, KyungHyun Cho, and Yoshua
  Bengio.
\newblock Empirical evaluation of gated recurrent neural networks on sequence
  modeling.
\newblock volume abs/1412.3555, 2014.

\bibitem{DauphinLanguage}
Yann~N. Dauphin, Angela Fan, Michael Auli, and David Grangier.
\newblock Language modeling with gated convolutional networks.
\newblock 2017.

\bibitem{DavePredictive}
Achal Dave, Olga Russakovsky, and Deva Ramanan.
\newblock Predictive-corrective networks for action detection.
\newblock 2017.

\bibitem{De2016Online}
Roeland De~Geest, Efstratios Gavves, Amir Ghodrati, Zhenyang Li, Cees Snoek,
  and Tinne Tuytelaars.
\newblock Online action detection.
\newblock In {\em CVPR}, 2016.

\bibitem{De}
Roeland De~Geest and Tinne Tuytelaars.
\newblock 2018 ieee winter conference on applications of computer vision (wacv)
  - modeling temporal structure with lstm for online action detection.
\newblock In {\em WACV}, 2018.

\bibitem{De2018}
Roeland De~Geest and Tinne Tuytelaars.
\newblock Modeling temporal structure with lstm for online action detection.
\newblock In {\em WACV}, pages 1549--1557, 2018.

\bibitem{DBLP:journals/corr/DonahueHGRVSD14}
Jeff Donahue, Lisa~Anne Hendricks, Sergio Guadarrama, Marcus Rohrbach,
  Subhashini Venugopalan, Kate Saenko, and Trevor Darrell.
\newblock Long-term recurrent convolutional networks for visual recognition and
  description.
\newblock In {\em CVPR}, volume abs/1411.4389, 2014.

\bibitem{Donahue2014Long}
Jeff Donahue, Lisa~Anne Hendricks, Marcus Rohrbach, Subhashini Venugopalan,
  Sergio Guadarrama, Kate Saenko, and Trevor Darrell.
\newblock Long-term recurrent convolutional networks for visual recognition and
  description.
\newblock 2014.

\bibitem{Donahue_2015_CVPR}
Jeffrey Donahue, Lisa Anne~Hendricks, Sergio Guadarrama, Marcus Rohrbach,
  Subhashini Venugopalan, Kate Saenko, and Trevor Darrell.
\newblock Long-term recurrent convolutional networks for visual recognition and
  description.
\newblock In {\em CVPR}, 2015.

\bibitem{Du2015Learning}
Tran Du, Lubomir Bourdev, Rob Fergus, Lorenzo Torresani, and Manohar Paluri.
\newblock Learning spatiotemporal features with 3d convolutional networks.
\newblock In {\em ICCV}, 2015.

\bibitem{DuA}
Tran Du, Heng Wang, Lorenzo Torresani, Jamie Ray, and Yann Lecun.
\newblock A closer look at spatiotemporal convolutions for action recognition.
\newblock In {\em CVPR}, 2018.

\bibitem{NIPS2019_8498}
Quanfu Fan, Chun-Fu~(Richard) Chen, Hilde Kuehne, Marco Pistoia, and David Cox.
\newblock More is less: Learning efficient video representations by big-little
  network and depthwise temporal aggregation.
\newblock In {\em NIPS}, pages 2261--2270,
  2019.

\bibitem{FeichtenhoferConvolutional}
Christoph Feichtenhofer, Axel Pinz, and Andrew Zisserman.
\newblock Convolutional two-stream network fusion for video action recognition.
\newblock In {\em CVPR}, 2016.

\bibitem{Furnari_2019_ICCV}
Antonino Furnari and Giovanni~Maria Farinella.
\newblock What would you expect? anticipating egocentric actions with
  rolling-unrolling lstms and modality attention.
\newblock In {\em ICCV}, 2019.

\bibitem{GammulleTwo}
Harshala Gammulle, Simon Denman, Sridha Sridharan, and Clinton Fookes.
\newblock Two stream lstm: A deep fusion framework for human action
  recognition.
\newblock 2017.

\bibitem{Gao_2018_ECCV}
Jiyang Gao, Kan Chen, and Ram Nevatia.
\newblock Ctap: Complementary temporal action proposal generation.
\newblock In {\em ECCV}, 2018.

\bibitem{Gao2018Revisiting}
Jiyang Gao and Ram Nevatia.
\newblock Revisiting temporal modeling for video-based person reid.
\newblock In {\em BMVC}, 2018.

\bibitem{DBLP:conf/bmvc/GaoYN17a}
Jiyang Gao, Zhenheng Yang, and Ram Nevatia.
\newblock {RED:} reinforced encoder-decoder networks for action anticipation.
\newblock In {\em BMVC}, 2017.

\bibitem{DBLP:journals/corr/abs-1903-09868}
Mingfei Gao, Mingze Xu, Larry~S. Davis, Richard Socher, and Caiming Xiong.
\newblock Startnet: Online detection of action start in untrimmed videos.
\newblock 2019.

\bibitem{Girdhar_2019_CVPR}
Rohit Girdhar, Joao Carreira, Carl Doersch, and Andrew Zisserman.
\newblock Video action transformer network.
\newblock In {\em CVPR}, 2019.

\bibitem{DBLP:conf/cvpr/GirshickDDM14}
Ross~B. Girshick, Jeff Donahue, Trevor Darrell, and Jitendra Malik.
\newblock Rich feature hierarchies for accurate object detection and semantic
  segmentation.
\newblock In {\em CVPR}, 2014.

\bibitem{DBLP:journals/corr/GkioxariM14}
Georgia Gkioxari and Jitendra Malik.
\newblock Finding action tubes.
\newblock {\em CoRR}, abs/1411.6031, 2014.

\bibitem{Graves1997Long}
Alex Graves.
\newblock Long short-term memory.
\newblock volume~9, pages 1735--1780, 1997.

\bibitem{DBLP:journals/corr/Graves13}
Alex Graves.
\newblock Generating sequences with recurrent neural networks.
\newblock {\em CoRR}, abs/1308.0850, 2013.

\bibitem{DBLP:conf/cvpr/GuSRVPLVTRSSM18}
Chunhui Gu, Chen Sun, David~A. Ross, Carl Vondrick, Caroline Pantofaru, Yeqing
  Li, Sudheendra Vijayanarasimhan, George Toderici, Susanna Ricco, Rahul
  Sukthankar, Cordelia Schmid, and Jitendra Malik.
\newblock {AVA:} {A} video dataset of spatio-temporally localized atomic visual
  actions.
\newblock In {\em CVPR}, 2018.

\bibitem{HaraCan}
Kensho Hara, Hirokatsu Kataoka, and Yutaka Satoh.
\newblock Can spatiotemporal 3d cnns retrace the history of 2d cnns and
  imagenet?
\newblock In {\em CVPR}, 2017.

\bibitem{DBLP:journals/corr/HeZRS15}
Kaiming He, Xiangyu Zhang, Shaoqing Ren, and Jian Sun.
\newblock Deep residual learning for image recognition.
\newblock volume abs/1512.03385, 2015.

\bibitem{Hoai2012Max}
Minh Hoai and Fernando De~La Torre.
\newblock Max-margin early event detectors.
\newblock In {\em CVPR}, 2012.

\bibitem{Hochreiter1997Long}
Sepp Hochreiter and Jürgen Schmidhuber.
\newblock Long short-term memory.
\newblock volume~9, pages 1735--1780, 1997.

\bibitem{DBLP:journals/neco/HochreiterS97}
Sepp Hochreiter and J{\"{u}}rgen Schmidhuber.
\newblock Long short-term memory.
\newblock {\em Neural Computation}, 1997.

\bibitem{Ioffe2015Batch}
Sergey Ioffe and Christian Szegedy.
\newblock Batch normalization: Accelerating deep network training by reducing
  internal covariate shift.
\newblock 2015.

\bibitem{Jhuang2013Towards}
Hueihan Jhuang, Juergen Gall, Silvia Zuffi, Cordelia Schmid, and Michael~J.
  Black.
\newblock Towards understanding action recognition.
\newblock In {\em ICCV}, 2013.

\bibitem{THUMOS14}
Y.-G. Jiang, J.~Liu, A.~Roshan~Zamir, G.~Toderici, I.~Laptev, M.~Shah, and
  R.~Sukthankar.
\newblock {THUMOS} challenge: Action recognition with a large number of
  classes.
\newblock 2014.

\bibitem{Jing2018Attention}
Xu~Jing, Zhao Rui, Zhu Feng, Huaming Wang, and Wanli Ouyang.
\newblock Attention-aware compositional network for person re-identification.
\newblock In {\em CVPR}, 2018.

\bibitem{DBLP:journals/corr/JozefowiczVSSW16}
Rafal J{\'{o}}zefowicz, Oriol Vinyals, Mike Schuster, Noam Shazeer, and Yonghui
  Wu.
\newblock Exploring the limits of language modeling.
\newblock volume abs/1602.02410, 2016.

\bibitem{KarAdaScan}
Amlan Kar, Nishant Rai, Karan Sikka, and Gaurav Sharma.
\newblock Adascan: Adaptive scan pooling in deep convolutional neural networks
  for human action recognition in videos.
\newblock 2017.

\bibitem{Karpathy2014Large}
Andrej Karpathy, George Toderici, Sanketh Shetty, Thomas Leung, and Fei~Fei Li.
\newblock Large-scale video classification with convolutional neural networks.
\newblock In {\em CVPR}, 2014.

\bibitem{Kay2017The}
Will Kay, Joao Carreira, Karen Simonyan, Brian Zhang, and Andrew Zisserman.
\newblock The kinetics human action video dataset.
\newblock 2017.

\bibitem{Ke2016Human}
Qiuhong Ke, Mohammed Bennamoun, Senjian An, Farid Boussaid, and Ferdous Sohel.
\newblock Human interaction prediction using deep temporal features.
\newblock In {\em ECCV}, 2016.

\bibitem{Ke_2019_CVPR}
Qiuhong Ke, Mario Fritz, and Bernt Schiele.
\newblock Time-conditioned action anticipation in one shot.
\newblock In {\em CVPR}, 2019.

\bibitem{8099873}
Y.~{Kong}, Z.~{Tao}, and Y.~{Fu}.
\newblock Deep sequential context networks for action prediction.
\newblock In {\em CVPR}, 2017.

\bibitem{KongInteractive}
Yu~Kong, Yunde Jia, and Yun Fu.
\newblock Interactive phrases: Semantic descriptionsfor human interaction
  recognition.
\newblock volume~36, pages 1775--1788, 2014.

\bibitem{Kong2014A}
Yu~Kong, Dmitry Kit, and Yun Fu.
\newblock A discriminative model with multiple temporal scales for action
  prediction.
\newblock In {\em ECCV}, 2014.

\bibitem{Kuehne2013HMDB51}
Hilde Kuehne, Hueihan Jhuang, Rainer Stiefelhagen, and Thomas Serre.
\newblock Hmdb51: A large video database for human motion recognition.
\newblock In {\em High Performance Computing in Science and Engineering ‘12},
  pages 571--582. Springer, 2013.

\bibitem{Laptev2008Learning}
Ivan Laptev, Marcin Marszalek, Cordelia Schmid, and Benjamin Rozenfeld.
\newblock Learning realistic human actions from movies.
\newblock In {\em CVPR}, 2008.

\bibitem{DBLP:journals/corr/LeaFVRH16}
Colin Lea, Michael~D. Flynn, Ren{\'{e}} Vidal, Austin Reiter, and Gregory~D.
  Hager.
\newblock Temporal convolutional networks for action segmentation and
  detection.
\newblock volume abs/1611.05267, 2016.

\bibitem{LiGlobal}
Jianing Li, Jingdong Wang, Qi~Tian, Wen Gao, and Shiliang Zhang.
\newblock Global-local temporal representations for video person
  re-identification.
\newblock 2019.

\bibitem{Li2014Prediction}
Kang Li and Yun Fu.
\newblock Prediction of human activity by discovering temporal sequence
  patterns.
\newblock volume~36, pages 1644--1657, 2014.

\bibitem{LiOnline}
Yanghao Li, Cuiling Lan, Junliang Xing, Wenjun Zeng, Chunfeng Yuan, and Jiaying
  Liu.
\newblock Online human action detection using joint classification-regression
  recurrent neural networks.
\newblock In {\em ECCV}, 2016.

\bibitem{DBLP:journals/corr/abs-1811-08383}
Ji~Lin, Chuang Gan, and Song Han.
\newblock Temporal shift module for efficient video understanding.
\newblock volume abs/1811.08383, 2018.

\bibitem{Lin_2018_ECCV}
Tianwei Lin, Xu~Zhao, Haisheng Su, Chongjing Wang, and Ming Yang.
\newblock Bsn: Boundary sensitive network for temporal action proposal
  generation.
\newblock In {\em ECCV}, 2018.

\bibitem{LinBSN}
Tianwei Lin, Xu~Zhao, Haisheng Su, Chongjing Wang, and Ming Yang.
\newblock Bsn: Boundary sensitive network for temporal action proposal
  generation.
\newblock In {\em ECCV}, 2018.

\bibitem{lin2017structured}
Zhouhan Lin, Minwei Feng, Cicero Nogueira~dos Santos, Mo~Yu, Bing Xiang, Bowen
  Zhou, and Yoshua Bengio.
\newblock A structured self-attentive sentence embedding.
\newblock 2017.

\bibitem{Liu2017Online}
Chunhui Liu, Yanghao Li, Yueyu Hu, and Jiaying Liu.
\newblock Online action detection and forecast via multitask deep recurrent
  neural networks.
\newblock In {\em ICASSP}, 2017.

\bibitem{Liu2018Multi}
Jiaying Liu, Yanghao Li, Sijie Song, Junliang Xing, and Wenjun Zeng.
\newblock Multi-modality multi-task recurrent neural network for online action
  detection.
\newblock 2018.

\bibitem{Liu2018Skeleton}
Jun Liu, Gang Wang, Ling~Yu Duan, Kamila Abdiyeva, and Alex~C. Kot.
\newblock Skeleton based human action recognition with global context-aware
  attention lstm networks.
\newblock volume~PP, pages 1--1, 2018.

\bibitem{Liu2017Global}
Jun Liu, Gang Wang, Ping Hu, Ling~Yu Duan, and Alex~C Kot.
\newblock Global context-aware attention lstm networks for 3d action
  recognition.
\newblock In {\em CVPR}, 2017.

\bibitem{DBLP:conf/eccv/LiuAESRFB16}
Wei Liu, Dragomir Anguelov, Dumitru Erhan, Christian Szegedy, Scott~E. Reed,
  Cheng{-}Yang Fu, and Alexander~C. Berg.
\newblock {SSD:} single shot multibox detector.
\newblock In {\em ECCV}, 2016.

\bibitem{Luong2015Effective}
Minh~Thang Luong, Hieu Pham, and Christopher~D. Manning.
\newblock Effective approaches to attention-based neural machine translation.
\newblock 2015.

\bibitem{Mahmud2017Joint}
Tahmida Mahmud, Mahmudul Hasan, and Amit~K. Roy-Chowdhury.
\newblock Joint prediction of activity labels and starting times in untrimmed
  videos.
\newblock In {\em CVPR}, 2017.

\bibitem{NgBeyond}
Yue~Hei Ng, Matthew Hausknecht, Sudheendra Vijayanarasimhan, Oriol Vinyals,
  Rajat Monga, and George Toderici.
\newblock Beyond short snippets: Deep networks for video classification.
\newblock In {\em CVPR}, 2015.

\bibitem{OordWaveNet}
Aaron Van~Den Oord, Sander Dieleman, Heiga Zen, Karen Simonyan, Oriol Vinyals,
  Alex Graves, Nal Kalchbrenner, Andrew Senior, and Koray Kavukcuoglu.
\newblock Wavenet: A generative model for raw audio.
\newblock 2016.

\bibitem{doi:10.5244/C.24.50:abbreviated}
Alonso Patron, Marcin Marszalek, Andrew Zisserman, and Ian Reid.
\newblock High five: Recognising human interactions in tv shows.
\newblock In {\em BMVC}, 2010.

\bibitem{DBLP:conf/eccv/PengS16}
Xiaojiang Peng and Cordelia Schmid.
\newblock Multi-region two-stream {R-CNN} for action detection.
\newblock In {\em ECCV}, 2016.

\bibitem{Qiu2017Learning}
Zhaofan Qiu, Ting Yao, and Mei Tao.
\newblock Learning spatio-temporal representation with pseudo-3d residual
  networks.
\newblock In {\em ICCV}, 2017.

\bibitem{Ren2017Faster}
S.~Ren, K.~He, R~Girshick, and J.~Sun.
\newblock Faster r-cnn: Towards real-time object detection with region proposal
  networks.
\newblock volume~39, pages 1137--1149, 2017.

\bibitem{Ren2015Faster}
Shaoqing Ren, Kaiming He, Ross Girshick, and Jian Sun.
\newblock Faster r-cnn: Towards real-time object detection with region proposal
  networks.
\newblock 2015.

\bibitem{Ryoo2012Human}
M.~S. Ryoo.
\newblock Human activity prediction: Early recognition of ongoing activities
  from streaming videos.
\newblock In {\em CVPR}, 2012.

\bibitem{Sharma2015Action}
Shikhar Sharma, Ryan Kiros, and Ruslan Salakhutdinov.
\newblock Action recognition using visual attention.
\newblock In {\em ICLR}, 2015.

\bibitem{Sharma2017Action}
Shikhar Sharma, Ryan Kiros, and Ruslan Salakhutdinov.
\newblock Action recognition using visual attention.
\newblock 2017.

\bibitem{ShouCDC}
Zheng Shou, Jonathan Chan, Alireza Zareian, Kazuyuki Miyazawa, and Shih~Fu
  Chang.
\newblock Cdc: Convolutional-de-convolutional networks for precise temporal
  action localization in untrimmed videos.
\newblock In {\em ICCV}, 2017.

\bibitem{Shou2018Online}
Zheng Shou, Junting Pan, Jonathan Chan, Kazuyuki Miyazawa, Hassan Mansour,
  Anthony Vetro, Xavier Giro-I-Nieto, and Shih~Fu Chang.
\newblock Online detection of action start in untrimmed, streaming videos.
\newblock In {\em CVPR}, 2018.

\bibitem{Shou2016Temporal}
Zheng Shou, Dongang Wang, and Shih~Fu Chang.
\newblock Temporal action localization in untrimmed videos via multi-stage
  cnns.
\newblock In {\em CVPR}, 2016.

\bibitem{Simonyan2014Two}
Karen Simonyan and Andrew Zisserman.
\newblock Two-stream convolutional networks for action recognition in videos.
\newblock In {\em NIPS}, 2014.

\bibitem{Simonyan2014Very}
Karen Simonyan and Andrew Zisserman.
\newblock Very deep convolutional networks for large-scale image recognition.
\newblock 2014.

\bibitem{Singh2016A}
Bharat Singh, Tim~K. Marks, Michael Jones, Oncel Tuzel, and Shao Ming.
\newblock A multi-stream bi-directional recurrent neural network for
  fine-grained action detection.
\newblock In {\em CVPR}, 2016.

\bibitem{singh2017online}
Gurkirt Singh, Suman Saha, Michael Sapienza, Philip~HS Torr, and Fabio
  Cuzzolin.
\newblock Online real-time multiple spatiotemporal action localisation and
  prediction.
\newblock In {\em ICCV}, pages 3637--3646, 2017.

\bibitem{Song2016An}
Sijie Song, Cuiling Lan, Junliang Xing, Wenjun Zeng, and Jiaying Liu.
\newblock An end-to-end spatio-temporal attention model for human action
  recognition from skeleton data.
\newblock In {\em AAAI}, 2016.

\bibitem{DBLP:conf/cvpr/SoomroIS16}
Khurram Soomro, Haroon Idrees, and Mubarak Shah.
\newblock Predicting the where and what of actors and actions through online
  action localization.
\newblock In {\em CVPR}, pages 2648--2657, 2016.

\bibitem{Soomro2012UCF101}
Khurram Soomro, Amir~Roshan Zamir, and Mubarak Shah.
\newblock Ucf101: A dataset of 101 human actions classes from videos in the
  wild.
\newblock 2012.

\bibitem{Srivastava2014Dropout}
Nitish Srivastava, Geoffrey Hinton, Alex Krizhevsky, Ilya Sutskever, and Ruslan
  Salakhutdinov.
\newblock Dropout: a simple way to prevent neural networks from overfitting.
\newblock volume~15, pages 1929--1958, 2014.

\bibitem{DBLP:journals/corr/SutskeverVL14}
Ilya Sutskever, Oriol Vinyals, and Quoc~V. Le.
\newblock Sequence to sequence learning with neural networks.
\newblock {\em CoRR}, abs/1409.3215, 2014.

\bibitem{DBLP:journals/corr/VaswaniSPUJGKP17}
Ashish Vaswani, Noam Shazeer, Niki Parmar, Jakob Uszkoreit, Llion Jones,
  Aidan~N. Gomez, Lukasz Kaiser, and Illia Polosukhin.
\newblock Attention is all you need.
\newblock volume abs/1706.03762, 2017.

\bibitem{DBLP:journals/corr/VondrickPT15}
Carl Vondrick, Hamed Pirsiavash, and Antonio Torralba.
\newblock Anticipating the future by watching unlabeled video.
\newblock volume abs/1504.08023, 2015.

\bibitem{inproceedings}
Carl Vondrick, Hamed Pirsiavash, and Antonio Torralba.
\newblock Anticipating visual representations from unlabeled video.
\newblock In {\em CVPR}, 2016.

\bibitem{Wang2013Dense}
Heng Wang, Alexander Kläser, Cordelia Schmid, and Cheng~Lin Liu.
\newblock Dense trajectories and motion boundary descriptors for action
  recognition.
\newblock 2013.

\bibitem{Wang2019Delving}
Hongsong Wang and Jiashi Feng.
\newblock Delving into 3d action anticipation from streaming videos.
\newblock 2019.

\bibitem{Wang2015Towards}
Limin Wang, Yuanjun Xiong, Wang Zhe, and Qiao Yu.
\newblock Towards good practices for very deep two-stream convnets.
\newblock 2015.

\bibitem{Wang2017Non}
Xiaolong Wang, Ross Girshick, Abhinav Gupta, and Kaiming He.
\newblock Non-local neural networks.
\newblock In {\em CVPR}, 2018.

\bibitem{Wu_2019_CVPR}
Chao-Yuan Wu, Christoph Feichtenhofer, Haoqi Fan, Kaiming He, Philipp
  Krahenbuhl, and Ross Girshick.
\newblock Long-term feature banks for detailed video understanding.
\newblock In {\em CVPR}, June 2019.

\bibitem{Wu2016Google}
Yonghui Wu, Mike Schuster, Zhifeng Chen, Quoc~V. Le, Mohammad Norouzi, Wolfgang
  Macherey, Maxim Krikun, Yuan Cao, Qin Gao, and Klaus Macherey.
\newblock Google's neural machine translation system: Bridging the gap between
  human and machine translation.
\newblock 2016.

\bibitem{DBLP:journals/corr/WuWJYX15}
Zuxuan Wu, Xi~Wang, Yu{-}Gang Jiang, Hao Ye, and Xiangyang Xue.
\newblock Modeling spatial-temporal clues in a hybrid deep learning framework
  for video classification.
\newblock volume abs/1504.01561, 2015.

\bibitem{DBLP:journals/corr/abs-1712-04851}
Saining Xie, Chen Sun, Jonathan Huang, Zhuowen Tu, and Kevin Murphy.
\newblock Rethinking spatiotemporal feature learning for video understanding.
\newblock volume abs/1712.04851, 2017.

\bibitem{Xiong2016CUHK}
Yuanjun Xiong, Limin Wang, Zhe Wang, Bowen Zhang, Hang Song, Wei Li, Dahua Lin,
  Yu~Qiao, Luc Van~Gool, and Xiaoou Tang.
\newblock Cuhk \& ethz \& siat submission to activitynet challenge 2016.
\newblock 2016.

\bibitem{XuR}
Huijuan Xu, Abir Das, and Kate Saenko.
\newblock R-c3d: Region convolutional 3d network for temporal activity
  detection.
\newblock In {\em ICCV}, 2017.

\bibitem{Xu_2019_ICCV}
Mingze Xu, Mingfei Gao, Yi-Ting Chen, Larry~S. Davis, and David~J. Crandall.
\newblock Temporal recurrent networks for online action detection.
\newblock In {\em ICCV}, 2019.

\bibitem{Yang2018Attention}
Fan Yang, Ke~Yan, Shijian Lu, Huizhu Jia, Xiaodong Xie, and Wen Gao.
\newblock Attention driven person re-identification.
\newblock pages S0031320318303133--, 2018.

\bibitem{yang2016hierarchical}
Zichao Yang, Diyi Yang, Chris Dyer, Xiaodong He, Alex Smola, and Eduard Hovy.
\newblock Hierarchical attention networks for document classification.
\newblock 2016.

\bibitem{Yeung2015Every}
Serena Yeung, Olga Russakovsky, Ning Jin, Mykhaylo Andriluka, Greg Mori, and
  Li~Fei-Fei.
\newblock Every moment counts: Dense detailed labeling of actions in complex
  videos.
\newblock 2015.

\bibitem{Yu2013Recognize}
Cao Yu, Daniel Barrett, Andrei Barbu, Siddharth Narayanaswamy, and Wang Song.
\newblock Recognize human activities from partially observed videos.
\newblock In {\em CVPR}, 2013.

\bibitem{Zeng_2019_ICCV}
Runhao Zeng, Wenbing Huang, Mingkui Tan, Yu~Rong, Peilin Zhao, Junzhou Huang,
  and Chuang Gan.
\newblock Graph convolutional networks for temporal action localization.
\newblock In {\em ICCV}, 2019.

\end{thebibliography}

\end{document}